\documentclass[aps,prl,amsmath,twocolumn,superscriptaddress,letterpaper,floatfix]{revtex4-1}
\usepackage{graphicx,color}
\usepackage{verbatim}
\usepackage{amssymb}   
\usepackage{amsmath}
\usepackage{amsfonts}
\usepackage{mathdots}
\usepackage{hyperref}
\usepackage{epsfig}

\usepackage{braket}
\usepackage{bm}
\begin{document}

\title{Pseudo-Ising superconductivity induced by $p$-wave magnetism}

\author{Zi-Ting Sun}
\affiliation{Department of Physics, The Hong Kong University of Science and Technology, Clear Water Bay, Hong Kong, China} 	
\author{Xilin Feng}
\affiliation{Department of Physics, The Hong Kong University of Science and Technology, Clear Water Bay, Hong Kong, China} 	
\author{Ying-Ming Xie} 
 \affiliation{RIKEN Center for Emergent Matter Science (CEMS), Wako, Saitama 351-0198, Japan} 	
 
\author{Benjamin T. Zhou} 
\affiliation{Thrust of Advanced Materials \& Quantum Science and Technology Center, The Hong Kong University of Science and Technology (Guangzhou), Nansha, Guangzhou, China}

\author{Jin-Xin Hu}\thanks{Contact author: jhuphy@ust.hk}
\affiliation{Department of Physics, The Hong Kong University of Science and Technology, Clear Water Bay, Hong Kong, China} 	

\author{K. T. Law}
\affiliation{Department of Physics, The Hong Kong University of Science and Technology, Clear Water Bay, Hong Kong, China}

	\date{\today}
	\begin{abstract}
Unconventional magnetic orders usually interplay with superconductivity in intriguing ways. Here we propose that a conventional superconductor in proximity to a compensated $p$-wave magnet exhibits behaviors analogous to those of Ising superconductivity found in transition-metal dichalcogenides, which we refer to as pseudo-Ising superconductivity. The pseudo-Ising superconductivity is characterized by several distinctive features: (i) it stays much more robust under strong $p$-wave magnetism than usual ferromagnetism or $d$-wave altermagnetism, thanks to the apparent time-reversal symmetry in $p$-wave spin splitting; (ii) in the low-temperature regime, a second-order superconducting phase transition occurs at a significantly enhanced in-plane upper critical magnetic field $B_{c2}$; (iii) the supercurrent-carrying state establishes non-vanishing out-of-plane spin magnetization, which is forbidden by symmetry in Rahsba and Ising superconductors. We further propose a spin-orbit-free scheme to realize Majorana zero modes by placing superconducting quantum wires on a $p$-wave magnet. Our work establishes a new form of unconventional superconductivity generated by $p$-wave magnetism.
	\end{abstract}
	\maketitle

\emph{Introduction.}---The interplay between superconductivity and magnetism is a fascinating subject in condensed matter physics. While usually thought to be competing against each other, superconductivity and magnetism can combine to produce exotic physics such as topological superconductivity~\cite{qi2010chiral,wang2015chiral,tokura2019magnetic,klinovaja2013topological,kezilebieke2020topological} and superconducting spin currents in superconductor-magnet hybrid structures~\cite{linder2015superconducting,maggiora2024superconductivity}. Moreover, the competition between spin magnetism and superconductivity is reconciled in systems with broken inversion symmetry, where spin-orbit couplings (SOCs) lift the spin degeneracy in normal-state electronic bands and generate mixing between spin-singlet and spin-triplet Cooper pairs in the superconducting state~\cite{yuan2014possible}. This leads to strong Pauli-limit violation~\cite{gor2001superconducting,frigeri2004superconductivity}, as exemplified by the Ising superconductivity in transition-metal dichalcogenides (TMDs) with strongly enhanced in-plane upper critical field $B_{c2}$~\cite{lu2015evidence,xi2016ising,de2018tuning,wickramaratne2020ising,sohn2018unusual,wang2019type,ilic2017enhancement,hsu2017topological,nakamura2017odd,zhou2016ising,he2018magnetic,xie2020strongly,xie2023orbital}.

 The recent discovery of altermagnetism~\cite{vsmejkal2022emerging,ma2021multifunctional,lee2024broken,mazin2022altermagnetism,vsmejkal2022giant,ghorashi2024altermagnetic,fang2024quantum,xiao2024spin,jiang2024enumeration,chen2024enumeration,lee2024fermi} has opened up new directions in the search for novel superconducting states brought about by unconventional magnetism. Many interesting phenomena have been uncovered in altermagnet-superconductor junctions or hybrid structures, such as unusual Andreev reflections~\cite{papaj2023andreev,sun2023andreev}, $\varphi$-Josephson junctions~\cite{ouassou2023dc,beenakker2023phase,lu2024varphi,yang2025topological}, finite-momentum Cooper pairing~\cite{zhang2024finite}, and nonlinear Edelstein effect~\cite{hu2025nonlinear,zyuzin2024magnetoelectric}. In contrast to the even-parity collinear altermagnet, $p$-wave magnet distinguishes itself with non-collinear spin orientations [Fig.~\ref{fig:fig1}(a)] and odd-parity momentum-space spin splitting [Fig.~\ref{fig:fig1}(b)]~\cite{hellenes2023exchange,brekke2024minimal,maeda2024theory,sukhachov2024impurity,chakraborty2024highly}. Possible superconducting phases under $p$-wave magnetism have garnered growing recent interest~\cite{maeda2024theory,kokkeler2024quantum,sukhachov2024coexistence}, while the rich implications of $p$-wave magnetism on superconductivity remain to be fully addressed.

\begin{figure}
		\centering
		\includegraphics[width=1\linewidth]{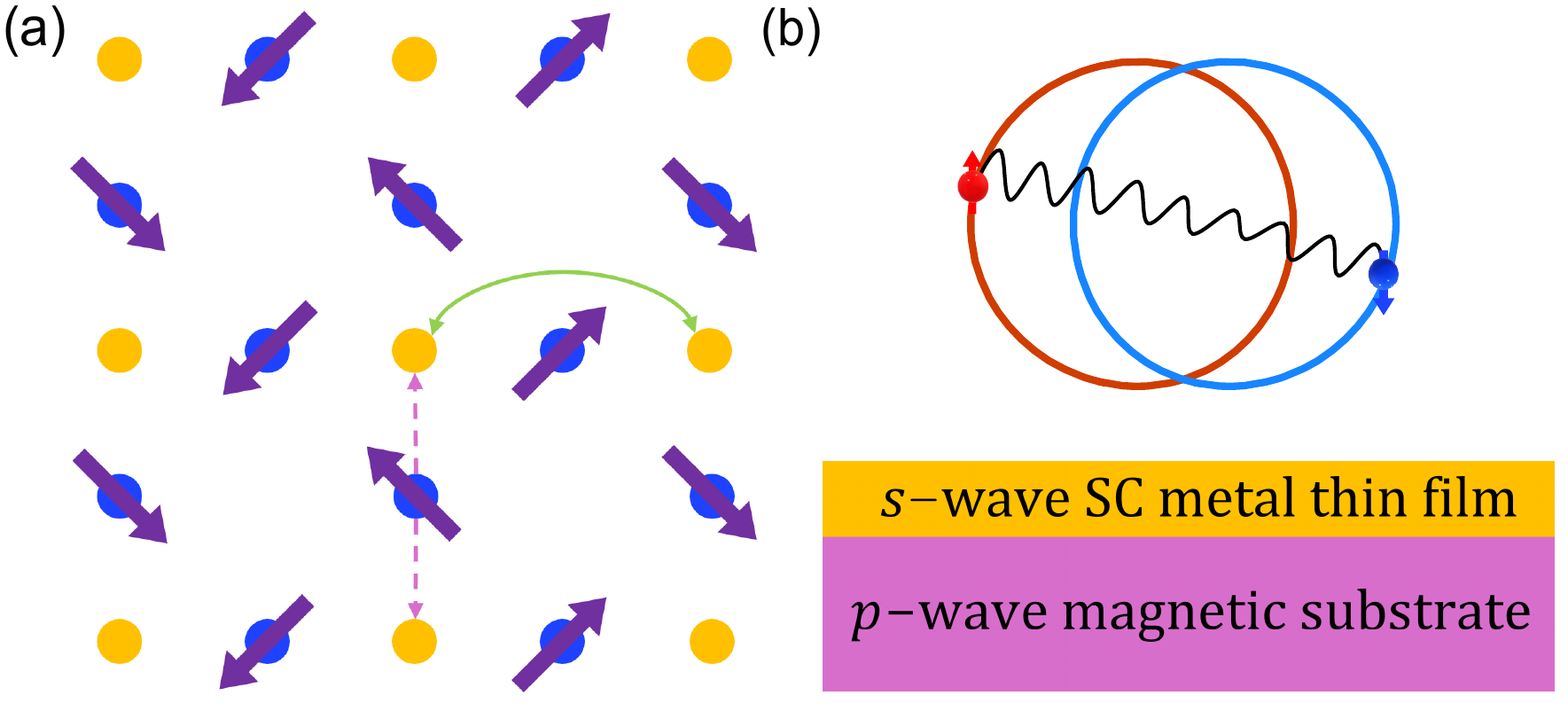}
		\caption{(a) Lattice model featuring a coplanar, non-collinear spin arrangement for compensated $p$-wave magnet, adapted from~\cite{hellenes2023exchange}. Nonmagnetic (magnetic) atoms are shown as orange (purple) sites. The green solid curve indicates spin-independent hopping, while the pink dashed line represents exchange-dependent hopping. (b) Upper panel: Fermi surfaces with $p$-wave spin splitting in momentum space, including the inter-pocket Cooper pairing in the superconducting state. Lower panel: A thin superconducting (SC) metal film placed atop a $p$-wave magnetic substrate to realize pseudo-Ising superconductivity via inverse proximity effect.}
		\label{fig:fig1}
\end{figure}
 In this Letter, we show that combining $p$-wave magnetism and superconductivity leads to a new form of superconductivity with its key features resembling Ising superconductivity in TMDs but in the complete absence of SOC. We refer to this new superconducting state as pseudo-Ising superconductivity. For concreteness, we consider a superconducting thin film placed on top of a $p$-wave magnet [Fig.~\ref{fig:fig1}(b)] which allows $p$-wave magnetism to be induced in the superconductor via inverse proximity effect~\cite{bergeret2004induced}. The superconducting state in this case not only survives under strong $p$-wave magnetism, but its in-plane upper critical field gets even enhanced by the presence of inverse proximity-induced $p$-wave magnetism, with the enhancement getting stronger upon increasing the strength of $p$-wave magnetism. Importantly, the superconductor acquires spin-triplet Cooper pairs through the $p$-wave magnetism and goes through a second-order superconductor-metal transition at $B_{c2}$. In the following, we discuss the mechanism behind pseudo-Ising superconductivity in detail and propose its potential applications in superconducting spintronics and the realization of Majorana zero modes (MZMs).

\vspace{2mm}
\emph{Effective Model.}---As depicted in Fig.~\ref{fig:fig1}(a), a $p$-wave magnet can be described by the model Hamiltonian $H(\boldsymbol{k})=H_0(\boldsymbol{k})+H_M(\boldsymbol{k})$ (material candidate CeNiAsO) adopted from~\cite{hellenes2023exchange,chakraborty2024highly} with $H_0$ the spin-independent hopping and $H_M$ the exchange-dependent hopping terms. 
The $H_M$ characterizes the $p$-wave spin splitting. To investigate the low-energy behaviors of the unconventional $p$-wave magnetism, we can derive the effective continuum model using standard perturbation theory  (see Supplemental Material (SM) I~\cite{NoteX}), which only contains the spin part for the hole-like top band around the $\Gamma$ pocket as 
\begin{equation}\label{eq:effmodel}
H_{eff}(\boldsymbol{k})=\xi(\boldsymbol{k})+\gamma_p(\boldsymbol{k})\sigma_z,
\end{equation}
where $\xi(\boldsymbol{k})=-t_y k_y^2-t_x k_x^2-\mu$ and
\begin{equation}
    \gamma_p(\boldsymbol{k})= J_p k_x(1-\frac{k_y^2}{2}),
    \label{eq:pmc}
\end{equation}
in which $\sigma$ are the spin Pauli matrices. For $p$-wave magnet proposed in Ref.~\cite{hellenes2023exchange,chakraborty2024highly}, the symmetries $\mathcal{T} \vec{t}$ and $\left[C_{2 \perp} \| \vec{t}\right]$ are preserved, where $\vec{t}$ is a translation by half a unit cell. In the momentum space, the former is an effective time-reversal symmetry rendering $\gamma_p(\boldsymbol{k})=-\gamma_p(-\boldsymbol{k})$ and the latter corresponds to spin symmetry $\left[C_{2 \perp} \| E\right]$, aligning spins in the out-of-plane orientation, i.e., only the $z$-component is nonvanishing. It is clear that the $p$-wave magnetism manifests itself as an odd-parity, out-of-plane non-relativistic spin-momentum locking. As we demonstrate below, the similarity between the $p$-wave magnetism and the Ising SOC \cite{lu2015evidence,xi2016ising} leads to what we call ``pseudo-Ising superconductivity" in a superconductor-$p$-wave magnet hybrid system.

\vspace{2mm}
\emph{Superconductivity \& $p$-wave magnetism.}---Based on the effective model, we examine the interplay between $p$-wave magnetism and superconductivity. As a natural illustration, we first introduce the model Hamiltonian of a two-dimensional $s$-wave superconductor thin film in proximity to a bulk magnetic substrate [e.g., $p$-wave magnet, see Fig.~\ref{fig:fig1}(b)], which reads
\begin{equation}
\mathcal{H}=  \sum_{\boldsymbol{k}, s, s^{\prime}} c_{\boldsymbol{k}, s}^{\dagger}h(\boldsymbol{k})_{ s, s^{\prime}} c_{\boldsymbol{k}, s^{\prime}} 
 -\frac{U}{A} \sum_{\boldsymbol{k}, \boldsymbol{k}^{\prime}} c_{\boldsymbol{k}, \uparrow}^{\dagger} c_{-\boldsymbol{k},\downarrow}^{\dagger} c_{-\boldsymbol{k}^{\prime}, \downarrow} c_{\boldsymbol{k}^{\prime}, \uparrow}
\end{equation}
where $c_{\boldsymbol{k}, s}^{\dagger}$ is the electronic creation operator with spin index $s=\uparrow / \downarrow$. $U$ is the attractive interaction establishing the $s$-wave superconducting order, and $A$ is the sample area. The hopping matrix $h(\boldsymbol{k})$ contains three parts: $h(\boldsymbol{k})=h_0(\boldsymbol{k})+h_J(\boldsymbol{k})+h_Z$, where $h_0(\boldsymbol{k})=-2 t\left(\cos k_x+\cos k_y\right)-\mu$, $h_J(\boldsymbol{k})=\gamma(\boldsymbol{k})\sigma_z$ refer to the magnetism induced by the inverse proximity effect from the magnetic substrate, and $h_Z=\boldsymbol{V}_{\parallel} \cdot \boldsymbol{\sigma}=\frac{1}{2} g_s \mu_{\mathrm{B}} \boldsymbol{B}_{\parallel} \cdot \boldsymbol{\sigma}$ is the Zeeman term. Here, $\boldsymbol{B}_{\parallel}=(B_x,B_y,0)$ denotes the in-plane magnetic field, and $g_s=2$ is the Landé $g$-factor. The orbital effect of the in-plane field is neglected, as usual, for a two-dimensional system.

 Within the mean-field approximation, the superconducting pairing amplitude is $\Delta=-U \sum_{\boldsymbol{k}}\left\langle c_{\boldsymbol{k}, \downarrow} c_{-\boldsymbol{k}, \uparrow}\right\rangle / A$. And the superconducting free energy density says
\begin{equation}\label{freen}
\mathcal{F}_s=\frac{|\Delta|^2}{U}- \sum_{k, n} \frac{1}{2 \beta A} \ln \left(1+e^{-\beta E_{k, n}}\right),
\end{equation}
with $E_{\boldsymbol{k}, n}$ the eigenvalues from the Bogoliubov-de Gennes (BdG) Hamiltonian $H_{\mathrm{BDG}}$, which reads
\begin{equation}\label{eq:bdg}
H_{\mathrm{BdG}}(\boldsymbol{k})=\left(\begin{array}{cc}
h(\boldsymbol{k}) & -i \Delta \sigma_y \\
i \Delta \sigma_y & -h^*(-\boldsymbol{k})
\end{array}\right).
\end{equation}
Then we introduce the magnetism brought by the inverse proximity effect from several typical magnets as
\begin{equation}
\label{eq:different_mag}
\gamma(\boldsymbol{k})\sigma_z=\left\{
\begin{aligned}
 &J\sigma_z \quad (s\text{-}\mathrm{wave}) \\   
  &J(\cos k_x-\cos k_y)\sigma_z/2 \quad (d\text{-}\mathrm{wave}) \\  
   &J\sin k_x \cos k_y\sigma_z \quad (p\text{-}\mathrm{wave}),
\end{aligned}
\right.
\end{equation}
which correspond to ferromagnetism ($s$-wave), $d$-wave altermagnetism~\cite{vsmejkal2022emerging}, and $p$-wave magnetism (discretization of Eq.~\eqref{eq:pmc}), respectively. For an easy comparison, the maximal spin splitting for all three cases is set to be $2J$, representing the strength of each type of magnetism. By solving the linearized gap equation:
\begin{equation}\label{lineareq}
    \left.\frac{\partial^2\mathcal{F}_s(\Delta,\boldsymbol{B}_{\parallel},T)}{\partial \Delta^2}\right |_{\Delta=0} =0,
\end{equation}
we obtain the superconducting critical temperature $T=T_c$ at $\boldsymbol{B}_{\parallel}=0$. In the absence of $J$, the critical temperature of the superconductor is $T_0$. Then, we plot $T_c$ as a function of $J$ for the three magnetic phases in Fig.~\ref{fig:fig2}(a). For the first two cases (ferromagnetism and $d$-wave altermagnetism), $T_c$ drops rapidly to zero as $J$ increases, whereas it remains robust even for large $p$-wave magnetism. This indicates that, unlike the other two cases, $p$-wave magnetism does not suppress $s$-wave superconductivity, indicating that the two orders can coexist. Although $p$-wave magnetism leaves $T_c$ (as well as the zero temperature pairing amplitude $\Delta_0\approx1.76k_B T_c$) unaffected, as we discuss next, it fundamentally alters the nature of superconducting pairing.

\vspace{2mm}
\begin{figure}
		\centering
        \hspace*{-0.2 cm}
		\includegraphics[width=1\linewidth]{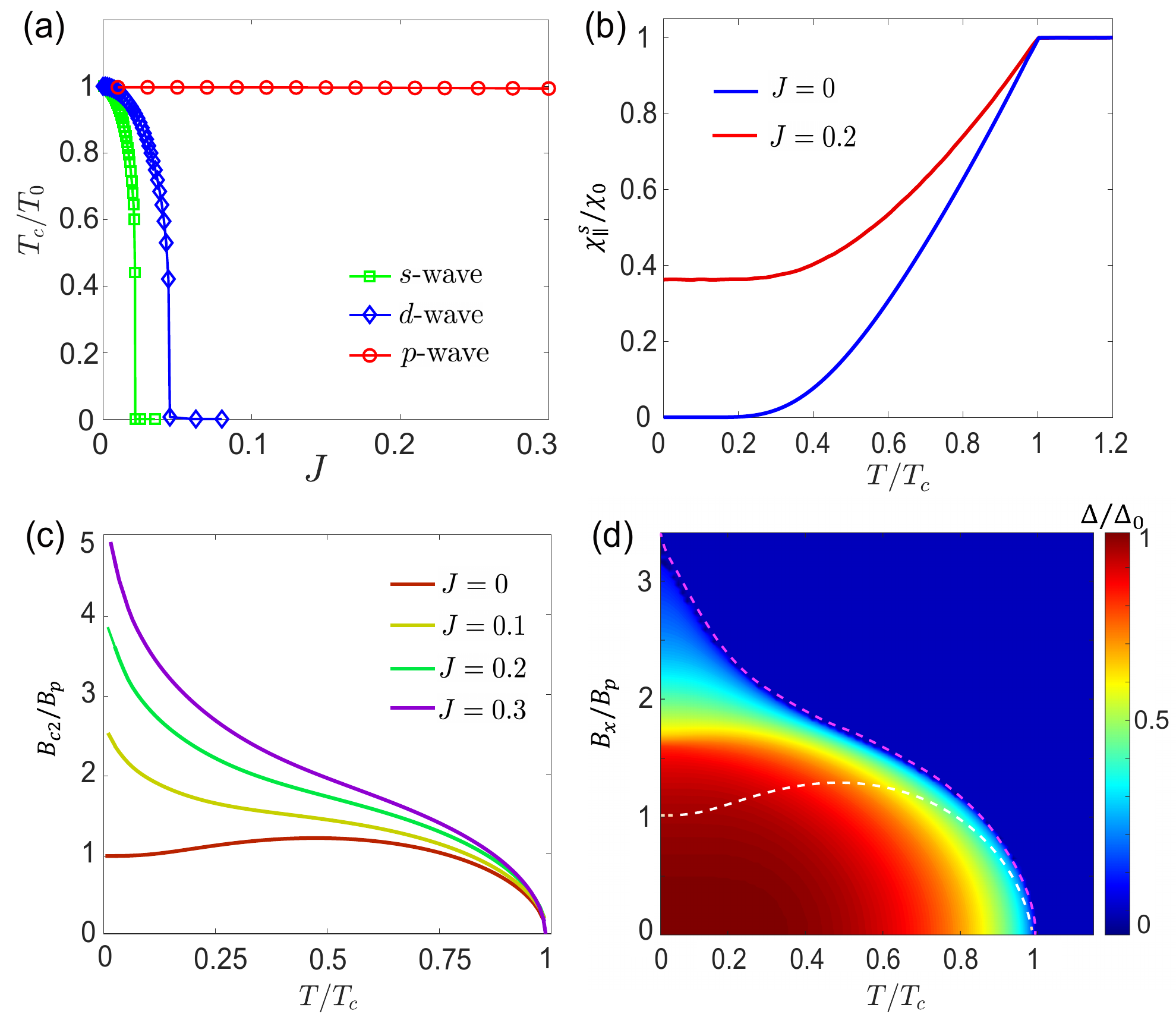}
		\caption{(a) Superconducting critical temperature with increasing strength for several types of magnetism. We label the $x$-axis as $J=J_s=J_p=J_d$ because they cause the same maximal band spin splitting. (b) In-plane spin susceptibility normalized by the normal-state value ($\chi^s_{\|}/\chi_{0}$) in the low-field limit as a function of temperature $T$ with (red line) and without (blue line) $p$-wave magnetism. (c) $B_{c2}\text{--}T$ curves for $\boldsymbol{B}_{\parallel}=B_x \hat{x}$ with different strengths of $p$-wave magnetism. (d) Pairing amplitude $\Delta$ as a function of both $B_x$ and $T$, when $J=0.2$. Other parameters: $t=1,U=1.5,\mu=-1$.}
		\label{fig:fig2}
\end{figure}
 
\emph{Pseudo-Ising superconductivity with $p$-wave magnetism.}---The unique nature of $p$-wave magnetism transforms a conventional superconductor into a pseudo-Ising superconductor. Here we identify two key characteristics of this state by analyzing its pairing correlations and spin susceptibility. With the anomalous Green function $F_{s s'}\left(\boldsymbol{k}, \tau_1; \boldsymbol{k}^{\prime}, \tau_2\right)=\left\langle T_\tau c_{\boldsymbol{k}, s}\left(\tau_1\right) c_{-\boldsymbol{k}^{\prime}, s'}\left(\tau_2\right)\right\rangle$ in hand, $F$ can be written as a compact matrix form in the frequency domain, yielding 
\begin{equation}
F\left(\boldsymbol{k}, i\omega_n\right)=\left[F_s\left(\boldsymbol{k}, i \omega_n\right)+\boldsymbol{F}_t\left(\boldsymbol{k}, i \omega_n\right) \cdot \boldsymbol{\sigma}\right] \Delta i \sigma_y,
\end{equation}
where $F_s$ and $\boldsymbol{F}_t$ parametrize the spin-singlet and triplet correlation functions, respectively, and $\omega_n=(2n+1)\pi k_{\text{B}}T$ is the fermionic Matsubara frequency.

By solving the Gor'kov equations, $F_s$ and $\boldsymbol{F}_t$ can be obtained as \cite{zhou2016ising,xie2020strongly}
\begin{equation}
\begin{gathered}\label{eq:paircorrelation}
F_s\left(\boldsymbol{k}, i \omega_n\right)=\frac{\varphi_{+}+\varphi_{-}}{2\varphi_{+}\varphi_{-}},\\
\boldsymbol{F}_t\left(\boldsymbol{k}, i \omega_n\right)=-2\frac{\xi(\boldsymbol{k})\gamma_p(\boldsymbol{k})\hat{z}+i \left[\omega_n + \gamma_p(\boldsymbol{k}) \hat{z} \times \right]\boldsymbol{V}_{\parallel}}{\varphi_{+}\varphi_{-}},
\end{gathered}
\end{equation}
with 
\begin{equation}\label{eq:phi}
\begin{split}
    \varphi_{ \pm}&\equiv\varphi_{ \pm}\left(\boldsymbol{k}, \boldsymbol{V}_{\parallel}, \Delta, \omega_n\right)\\
    &=\Delta^2+\omega_n^2- V^2_{\parallel}+\gamma_p^2(\boldsymbol{k})+\xi^2(\boldsymbol{k})\\
    &\pm2\left[\gamma_p^2(\boldsymbol{k}) \xi^2(\boldsymbol{k})-V^2_{\parallel} \left(\omega_n^2+\gamma_p^2(\boldsymbol{k})\right)\right]^{1 / 2},
\end{split}
\end{equation}
and $V_{\parallel}=|\boldsymbol{V}_{\parallel}|$. Notably, the presence of $p$-wave magnetism produces a nonvanishing {\it out-of-plane} antiparallel spin-triplet pairing correlation with $p$-wave symmetry (the first term in the numerator of $\boldsymbol{F}_t$ in Eq.~\eqref{eq:paircorrelation}, which is $\propto \gamma_p(\boldsymbol{k})\hat{z}$) despite the mean field pairing potential being $s$-wave. This reveals an unusual connection between unconventional magnetism and unconventional superconductivity, and suggests the emergence of topological superconductivity mediated by $p$-wave magnetism: in the basis of in-plane spins, these triplet Cooper pairs have their spinor parts given by equal-spin configurations as in Ising superconductors~\cite{zhou2016ising}. When subjected to an in-plane magnetic field $\boldsymbol{B}_{\parallel}$, these in-plane equal-spin Cooper pairs can align their spin magnetic moments along the magnetic field direction to save spin magnetization energy.

The in-plane equal-spin pairing correlation also endows the pseudo-Ising superconductivity with a finite spin susceptibility $\chi^s_{\|}$ at zero temperature, which can be evaluated by~\cite{frigeri2004spin}
\begin{equation}
\chi^s_{\|} = -\frac{1}{2}\mu_B^2 k_B T  \sum_{\boldsymbol{k},\omega_n} \text{Tr}\left [ \eta_x \mathcal{G}(\boldsymbol{k}, \omega_n) \eta_x \mathcal{G}(\boldsymbol{k}, \omega_n) \right ],
\end{equation}
where $\eta_a=\mathrm{diag}(\sigma_a,-\sigma_a^*)$ and $\mathcal{G}(\bm{k}, \omega_n)=[i\omega_n-H_{\mathrm{BdG}}]^{-1}$ is the Gor'kov Green's function. The normal-state spin susceptibility $\chi_0$ is directly given by $\chi^s_{\parallel}$($\Delta = 0$). In Fig.~\ref{fig:fig2}(b), $\chi^s_{\parallel}/\chi_0$ is plotted with and without $p$-wave magnetism. Consequently, the superconducting spin susceptibility $\chi^s_{\parallel}$ does not vanish at $T=0$, since the superconducting free energy density gets a compensation by $\bm{B}_{\parallel}$ as $-\chi^s_{\parallel}\bm{B}_{\parallel}^2/2$. This suggests enhancement in the in-plane $B_{c2}$~\cite{xie2020spin}.

\emph{Enhanced in-plane upper critical fields.}---To further elucidate the nature of pseudo-Ising superconductivity, we illustrate the relation between the in-plane upper critical fields $B_{c2}$ and temperature $T$ by solving the linearized gap equation Eq.~\eqref{lineareq} with $(B_x,B_y)=(B_{c2},0)$. The $B_{c2}$ as a function of temperature $T$ with different strengths of $p$-wave magnetism ($J$) are shown in Fig.~\ref{fig:fig2}(c). In the absence of $J$, superconductivity is destroyed by the paramagnetic effect at the Pauli limit $B_p \approx \Delta_0 / \sqrt{2} \mu_B$ near zero temperature (brown curve in Fig.~\ref{fig:fig2}(c) and white dashed curve in Fig.~\ref{fig:fig2}(d)). However, under strong $p$-wave magnetism, the $B_{c2}$ becomes significantly enhanced, reaching values several times the Pauli limit [other three curves in Fig.~\ref{fig:fig2}(c)]. 

In particular, in conventional superconductors, the low-temperature regions of the $B_{c2}\text{--}T$ curves derived from linearized gap equations ($0<T<T_1 \approx 0.5 T_c$) represent the supercooling critical field rather than the true $B_{c2}$ \cite{maki1964pauli}. The superconductor-metal transition at $B_{c2}$ exhibits a first-order nature within this regime. Our pseudo-Ising superconductor with $p$-wave magnetism, in contrast, exhibits distinct behaviors: by minimizing the free energy density $\mathcal{F}s$ with respect to $\Delta$ as a function of the magnetic field $B_{x}$ and temperature $T$ [Fig.~\ref{fig:fig2}(d)], we observe that the superconducting gap $\Delta$ vanishes continuously with increasing $B_x$ in the low temperature regime—a hallmark of second-order phase transitions at $B_{c2}$. This behavior is reminiscent of the continuous superconductor-metal transition observed in Ising superconductors~\cite{sohn2018unusual}. We note by passing that the analysis for pseudo-Ising superconductivity discussed above also applies to higher angular momentum odd-parity magnets such as $f$-wave magnet (see SM III~\cite{NoteX}).

\begin{figure}
		\centering
        \hspace*{-0.3 cm}
		\includegraphics[width=1\linewidth]{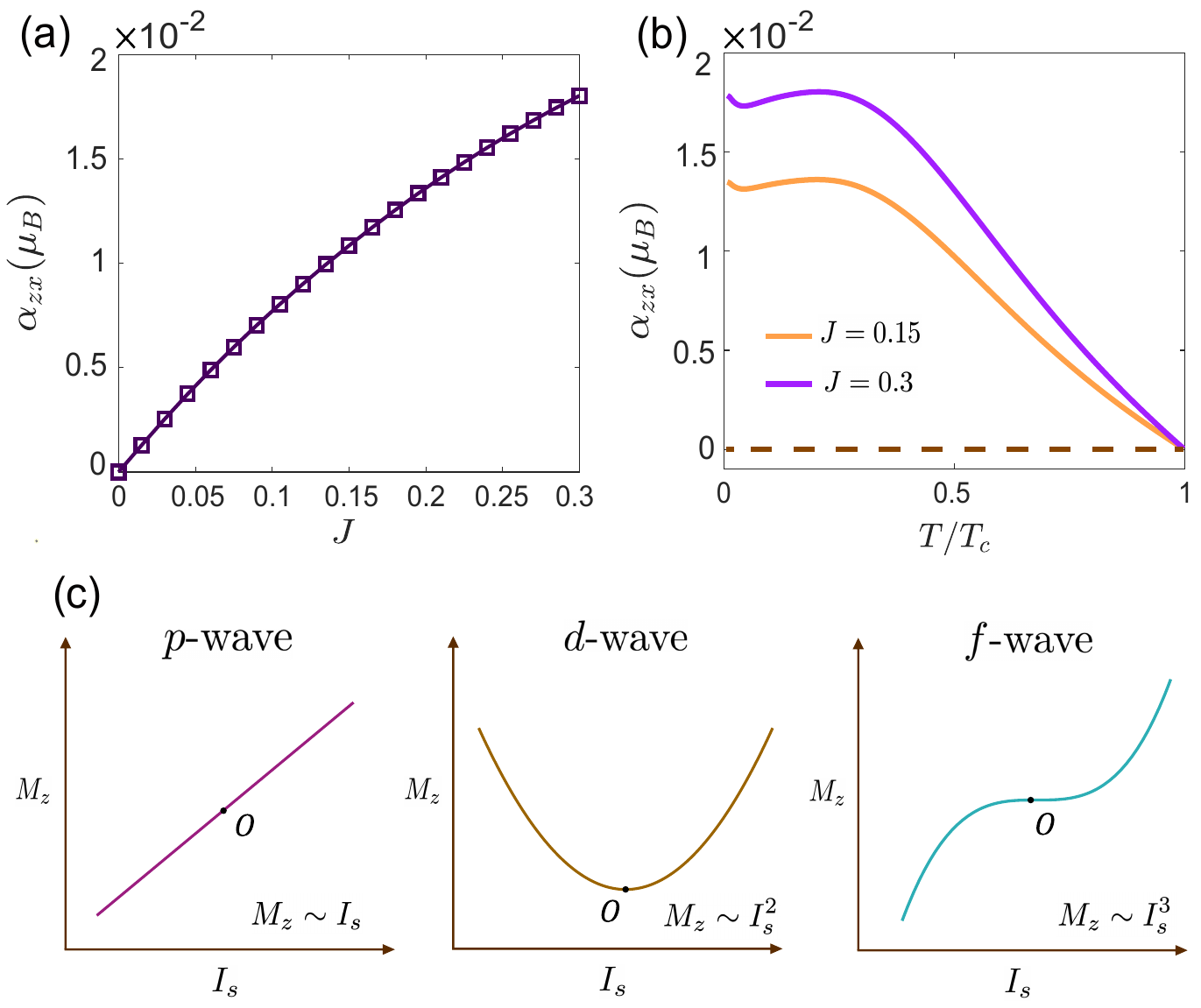}
		\caption{(a) The linear vertical Edelstein susceptibility $\alpha_{zx}$ as a function of the strength $J$ of $p$-wave magnetism at $T=0.2T_c$. (b) $\alpha_{zx}$ as a function of the temperature $T$. We also take a small Rashba SOC with $\lambda=0.03$. We adopt the standard BCS temperature dependence as $\Delta(T) = \Delta_0 \mathrm{tanh}(1.74\sqrt{T_c/T - 1})$ (for $p$-wave magnetism, this is verified numerically). Parameters:  $(\Delta_0, \mu) = (0.06,-1)$. (c) Linear and nonlinear vertical Edelstein effects induced by $p$, $d$, $f$ magnetic orders.}
		\label{fig:fig3}
\end{figure}

\emph{Dominant vertical Edelstein effect.}---In noncentrosymmetric superconductors with Rashba SOC, the supercurrent-induced magnetization—known as the Rashba-Edelstein effect~\cite{edelstein1995magnetoelectric,edelstein2005magnetoelectric}—is confined to the in-plane direction. In contrast, we show a qualitatively distinct behavior: $p$-wave magnetism produces a robust out-of-plane spin magnetization when an in-plane supercurrent is applied. Concretely, the Edelstein effect is expressed as $\delta M_a=\alpha_{ab} q_b$, where the magneto-electric susceptibility $\alpha_{ab}$ reads
\begin{equation}
\alpha_{ab}=-\frac{g_s\mu_B}{8\beta }\sum_{\boldsymbol{k},\omega_n}\mathrm{Tr}[\eta_a \mathcal{G}(\boldsymbol{k}, \omega_n) \hat{v}_b \mathcal{G}(\boldsymbol{k}, \omega_n)], 
\label{eq:linear_sus}
\end{equation}
where $\hat{v}_a=\mathrm{diag}[\mathcal{V}_a(\bm{k}),-\mathcal{V}_a^*(-\bm{k})]$ with the velocity operator $\mathcal{V}_a(\bm{k})=\partial h(\bm{k})/\partial k_a$. 

\begin{figure}[h]
		\centering
		\includegraphics[width=1\linewidth]{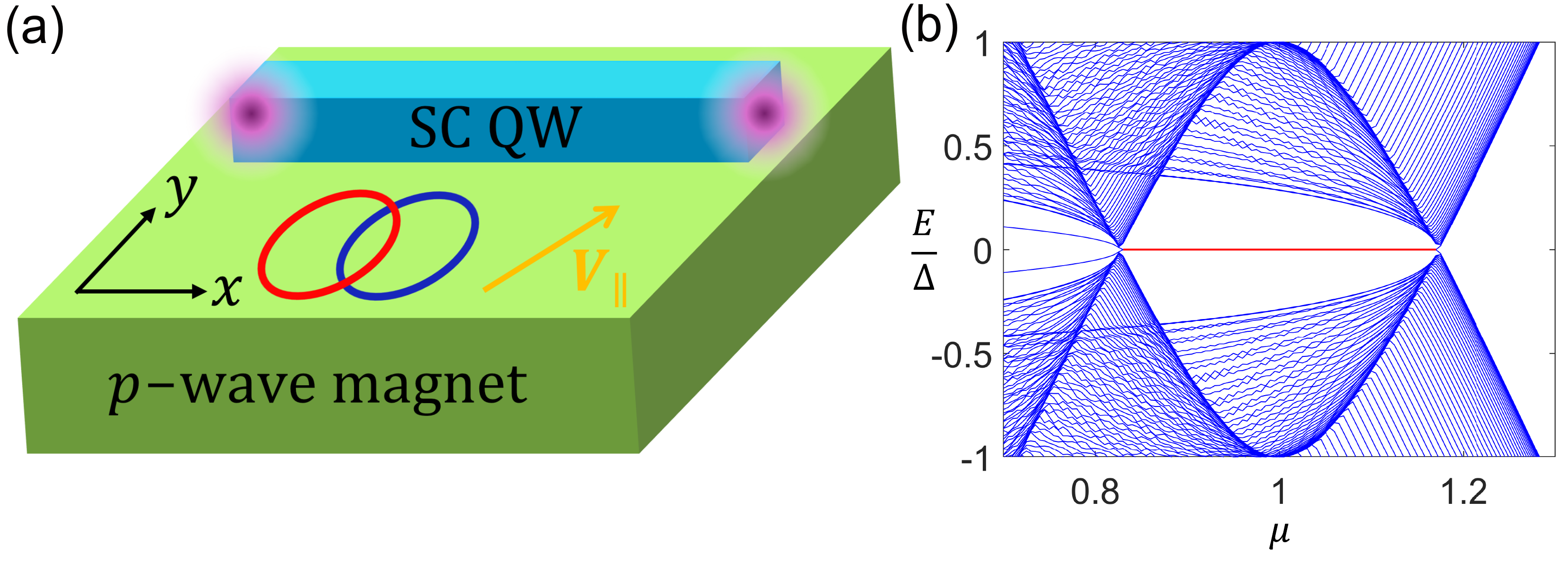}
		\caption{(a) A superconducting QW is placed on top of a $p$-wave magnetic substrate. MZMs (purple dots) appear at the ends of QW when an in-plane magnetic field $\boldsymbol{V}_{\parallel}$ is applied. (b) The energy spectrum of the setup in (a) as a function of the chemical potential of the wire, using the tight-binding model in Eq.~\eqref{majorana}. The red line highlights the topological regime with MZMs. Here we only plot the $\Delta^2+(2t_x-\mu)^2<V_y^2$ branch as a representative. Parameters: $J_p=-0.1$, $t_x=0.5$, $\Delta=0.1$, $V_y=0.2$.}
		\label{fig:fig4}
\end{figure}

In Fig.~\ref{fig:fig3}(a) we present the calculated vertical magneto-electric susceptibility $\alpha_{zx}$ as a function of $p$-wave magnetism strength $J$. In Fig.~\ref{fig:fig3}(b), we also show the temperature dependence of $\alpha_{zx}$ derived from Eq.~\eqref{eq:linear_sus}. Due to the $\left[C_{2 \perp} \| E\right]$ symmetry in momentum space, $\alpha_{zy}$ vanishes, leaving $\alpha_{zx}$ as the only nonzero term. It is anticipated that $\alpha_{zx}$ approaches zero near $T_c$ due to the fact that the spin polarization is associated with the supercurrent. Crucially, we note that at zero temperature, $\alpha_{zx}$ should vanish since $\langle s_z \rangle$ is conserved. Therefore, in our calculations, we include a small Rashba SOC term $\lambda(\sin k_x \sigma_y-\sin k_y \sigma_x)$ arising from the interface effect in the hybrid structure. The $\langle s_z \rangle$ conservation is specific to our model Hamiltonian~\eqref{eq:different_mag} and is not a general feature in realistic $p$-wave magnetic materials (more details, see SM IV~\cite{NoteX}). We emphasize that the vertical Edelstein effect is a result of the momentum-dependent $p$-wave magnetism, which pins the spin orientation to the out-of-plane direction. Interestingly, unconventional magnetic orders delineate various types of magneto-electric responses. As depicted in Fig.~\ref{fig:fig3}(c), the $p$, $d$, $f$-wave magnetic orders give rise to the first, second, and third order vertical Edelstein effects as the leading order effects, respectively. These nonlinear Edelstein effects stem from extra symmetry constraints. The case of $d$-wave has been reported recently~\cite{hu2025nonlinear,zyuzin2024magnetoelectric}, while the case of $f$-wave can be found in detail in SM III~\cite{NoteX}. Experimentally, a superconducting quantum interference device (SQUID) can be used to probe the magnetic flux change induced by the supercurrent~\cite{chirolli2022colossal}.

\emph{Possible route to MZMs utilizing $p$-wave magnetism.}---Finally we propose a feasible route to realize the MZMs using $p$-wave magnetism without any relativistic SOC. Considering a 1D superconducting quantum wire (QW) placed on the surface of a thick $p$-wave magnet [see Fig.~\ref{fig:fig4}(a)], the BdG Hamiltonian can be written as
\begin{equation}\label{majorana}
h_{\mathrm{BdG}}=E_{k} \rho_z+J_p \sin k_x \sigma_z+V_y \sigma_y+\Delta \rho_y \sigma_y,
\end{equation}
where $E_k=\left(2 t_x \cos k_x-\mu\right)$. Here $\rho$ and $\sigma$ are Pauli matrices in particle-hole and spin space. $J_p$ and $\Delta$ denote the proximity-induced $p$-wave magnetism and superconducting pairing, respectively. We assume the QW aligns with the $p$-wave spin splitting in the $x$-direction, and stays narrow vertically. If a misalignment angle $\alpha$ exists between them, $J_p$ will be renormalized as $J_p \cos \alpha$, which only introduces minor corrections to the topological regime. We take the in-plane Zeeman field $\boldsymbol{V}_{\parallel}=(0,V_y)$ for simplicity. The Hamiltonian is in class D with a $\mathbb{Z}_2$ classification, which is topological when an odd number of partially filled subbands are filled \cite{alicea2012new,beenakker2013search,elliott2015colloquium,lutchyn2018majorana,prada2020andreev,flensberg2021engineered,yazdani2023hunting}. The topological parameter regime is given by $\Delta^2+(2t_x\pm\mu)^2<V_y^2$, where the QW under open boundary condition hosts MZMs as end states, shown in Fig.~\ref{fig:fig4}(b). This proposed scheme based on pseudo-Ising superconductivity presents two key advantages over the state-of-the-art engineering of MZMs: First, light-element superconducting QWs can be much cleaner than Rashba nanowires, thus mitigating the effects of disorder~\cite{liu2012zero}. Second, the robustness of pseudo-Ising superconductivity against in-plane Zeeman fields allows the application of strong in-plane fields, which significantly extends the topological parameter regimes for Zeeman fields that would otherwise destroy superconducting pairing in usual Majorana nanowires \cite{xie2020strongly}.

\emph{Conclusion and Discussions.}---Using a low-energy effective model, we illustrate the mechanism of pseudo-Ising superconductivity in the hybrid structure formed by a conventional superconductor and a $p$-wave magnet. The pseudo-Ising superconducting state is similar to Ising superconductivity in terms of their pairing correlations and the nature of field-driven metal-superconductor transitions, while differing in its magneto-electric and topological properties due to its unique $p$-wave symmetry. Notably, while our two-band model in Eq.~\eqref{eq:effmodel} is obtained by integrating out degrees of freedom far away from the Fermi surface, it captures all the essential physics of pseudo-Ising superconductivity. We give further support to this claim by more realistic calculations based on the full tight binding models~\cite{hellenes2023exchange,brekke2024minimal}, see SM IV~\cite{NoteX}.

For the proposed mechanism of pseudo-Ising superconductivity to work, the $p$-wave magnetic order should remain robust in the presence of an in-plane magnetic field. On the other hand, we note that in some non-collinear antiferromagnets, a strong magnetic field may rearrange the spin orientations and possibly drive magnetic phase transitions. Nevertheless, our analysis is expected to hold as long as the applied in-plane field is lower than the critical field for magnetic phase transitions. In particular, for non-collinear magnets the critical magnetic field can be as high as $20$ Teslas~\cite{prokevs2020noncollinear,shi2021robust}, which goes far beyond the range of in-plane fields considered in our work (e.g., $B_{c2}\approx 2.5$ T for Al thin film near zero temperature~\cite{wu2019induced}). Thus, we believe the prerequisites of robust $p$-wave magnetism can be met by a suitable choice of $p$-wave magnet and superconducting materials, and the pseudo-Ising superconductivity can be realized under realistic conditions.

\emph{Acknowledgements}---We thank Xun-Jiang Luo and Akito Daido for inspiring discussions. J.X.H and K.T.L. acknowledge the support of the Ministry of Science and Technology, China, and Hong Kong Research Grant Council through Grants No. 2020YFA0309600, No. RFS2021-6S03, No. C6053-23G, No. AoE/P-701/20, No. 16310520, No. 16310219, No. 16307622, and No. 16311424.

\bibliographystyle{apsrev4-1}
\bibliography{main}

\begin{thebibliography}{69}%
\makeatletter
\providecommand \@ifxundefined [1]{%
 \@ifx{#1\undefined}
}%
\providecommand \@ifnum [1]{%
 \ifnum #1\expandafter \@firstoftwo
 \else \expandafter \@secondoftwo
 \fi
}%
\providecommand \@ifx [1]{%
 \ifx #1\expandafter \@firstoftwo
 \else \expandafter \@secondoftwo
 \fi
}%
\providecommand \natexlab [1]{#1}%
\providecommand \enquote  [1]{``#1''}%
\providecommand \bibnamefont  [1]{#1}%
\providecommand \bibfnamefont [1]{#1}%
\providecommand \citenamefont [1]{#1}%
\providecommand \href@noop [0]{\@secondoftwo}%
\providecommand \href [0]{\begingroup \@sanitize@url \@href}%
\providecommand \@href[1]{\@@startlink{#1}\@@href}%
\providecommand \@@href[1]{\endgroup#1\@@endlink}%
\providecommand \@sanitize@url [0]{\catcode `\\12\catcode `\$12\catcode `\&12\catcode `\#12\catcode `\^12\catcode `\_12\catcode `\%12\relax}%
\providecommand \@@startlink[1]{}%
\providecommand \@@endlink[0]{}%
\providecommand \url  [0]{\begingroup\@sanitize@url \@url }%
\providecommand \@url [1]{\endgroup\@href {#1}{\urlprefix }}%
\providecommand \urlprefix  [0]{URL }%
\providecommand \Eprint [0]{\href }%
\providecommand \doibase [0]{http://dx.doi.org/}%
\providecommand \selectlanguage [0]{\@gobble}%
\providecommand \bibinfo  [0]{\@secondoftwo}%
\providecommand \bibfield  [0]{\@secondoftwo}%
\providecommand \translation [1]{[#1]}%
\providecommand \BibitemOpen [0]{}%
\providecommand \bibitemStop [0]{}%
\providecommand \bibitemNoStop [0]{.\EOS\space}%
\providecommand \EOS [0]{\spacefactor3000\relax}%
\providecommand \BibitemShut  [1]{\csname bibitem#1\endcsname}%
\let\auto@bib@innerbib\@empty
\bibitem [{\citenamefont {Qi}\ \emph {et~al.}(2010)\citenamefont {Qi}, \citenamefont {Hughes},\ and\ \citenamefont {Zhang}}]{qi2010chiral}%
  \BibitemOpen
  \bibfield  {author} {\bibinfo {author} {\bibfnamefont {X.-L.}\ \bibnamefont {Qi}}, \bibinfo {author} {\bibfnamefont {T.~L.}\ \bibnamefont {Hughes}}, \ and\ \bibinfo {author} {\bibfnamefont {S.-C.}\ \bibnamefont {Zhang}},\ }\href@noop {} {\bibfield  {journal} {\bibinfo  {journal} {Physical Review B—Condensed Matter and Materials Physics}\ }\textbf {\bibinfo {volume} {82}},\ \bibinfo {pages} {184516} (\bibinfo {year} {2010})}\BibitemShut {NoStop}%
\bibitem [{\citenamefont {Wang}\ \emph {et~al.}(2015)\citenamefont {Wang}, \citenamefont {Zhou}, \citenamefont {Lian},\ and\ \citenamefont {Zhang}}]{wang2015chiral}%
  \BibitemOpen
  \bibfield  {author} {\bibinfo {author} {\bibfnamefont {J.}~\bibnamefont {Wang}}, \bibinfo {author} {\bibfnamefont {Q.}~\bibnamefont {Zhou}}, \bibinfo {author} {\bibfnamefont {B.}~\bibnamefont {Lian}}, \ and\ \bibinfo {author} {\bibfnamefont {S.-C.}\ \bibnamefont {Zhang}},\ }\href@noop {} {\bibfield  {journal} {\bibinfo  {journal} {Physical Review B}\ }\textbf {\bibinfo {volume} {92}},\ \bibinfo {pages} {064520} (\bibinfo {year} {2015})}\BibitemShut {NoStop}%
\bibitem [{\citenamefont {Tokura}\ \emph {et~al.}(2019)\citenamefont {Tokura}, \citenamefont {Yasuda},\ and\ \citenamefont {Tsukazaki}}]{tokura2019magnetic}%
  \BibitemOpen
  \bibfield  {author} {\bibinfo {author} {\bibfnamefont {Y.}~\bibnamefont {Tokura}}, \bibinfo {author} {\bibfnamefont {K.}~\bibnamefont {Yasuda}}, \ and\ \bibinfo {author} {\bibfnamefont {A.}~\bibnamefont {Tsukazaki}},\ }\href@noop {} {\bibfield  {journal} {\bibinfo  {journal} {Nature Reviews Physics}\ }\textbf {\bibinfo {volume} {1}},\ \bibinfo {pages} {126} (\bibinfo {year} {2019})}\BibitemShut {NoStop}%
\bibitem [{\citenamefont {Klinovaja}\ \emph {et~al.}(2013)\citenamefont {Klinovaja}, \citenamefont {Stano}, \citenamefont {Yazdani},\ and\ \citenamefont {Loss}}]{klinovaja2013topological}%
  \BibitemOpen
  \bibfield  {author} {\bibinfo {author} {\bibfnamefont {J.}~\bibnamefont {Klinovaja}}, \bibinfo {author} {\bibfnamefont {P.}~\bibnamefont {Stano}}, \bibinfo {author} {\bibfnamefont {A.}~\bibnamefont {Yazdani}}, \ and\ \bibinfo {author} {\bibfnamefont {D.}~\bibnamefont {Loss}},\ }\href@noop {} {\bibfield  {journal} {\bibinfo  {journal} {Physical review letters}\ }\textbf {\bibinfo {volume} {111}},\ \bibinfo {pages} {186805} (\bibinfo {year} {2013})}\BibitemShut {NoStop}%
\bibitem [{\citenamefont {Kezilebieke}\ \emph {et~al.}(2020)\citenamefont {Kezilebieke}, \citenamefont {Huda}, \citenamefont {Va{\v{n}}o}, \citenamefont {Aapro}, \citenamefont {Ganguli}, \citenamefont {Silveira}, \citenamefont {G{\l}odzik}, \citenamefont {Foster}, \citenamefont {Ojanen},\ and\ \citenamefont {Liljeroth}}]{kezilebieke2020topological}%
  \BibitemOpen
  \bibfield  {author} {\bibinfo {author} {\bibfnamefont {S.}~\bibnamefont {Kezilebieke}}, \bibinfo {author} {\bibfnamefont {M.~N.}\ \bibnamefont {Huda}}, \bibinfo {author} {\bibfnamefont {V.}~\bibnamefont {Va{\v{n}}o}}, \bibinfo {author} {\bibfnamefont {M.}~\bibnamefont {Aapro}}, \bibinfo {author} {\bibfnamefont {S.~C.}\ \bibnamefont {Ganguli}}, \bibinfo {author} {\bibfnamefont {O.~J.}\ \bibnamefont {Silveira}}, \bibinfo {author} {\bibfnamefont {S.}~\bibnamefont {G{\l}odzik}}, \bibinfo {author} {\bibfnamefont {A.~S.}\ \bibnamefont {Foster}}, \bibinfo {author} {\bibfnamefont {T.}~\bibnamefont {Ojanen}}, \ and\ \bibinfo {author} {\bibfnamefont {P.}~\bibnamefont {Liljeroth}},\ }\href@noop {} {\bibfield  {journal} {\bibinfo  {journal} {Nature}\ }\textbf {\bibinfo {volume} {588}},\ \bibinfo {pages} {424} (\bibinfo {year} {2020})}\BibitemShut {NoStop}%
\bibitem [{\citenamefont {Linder}\ and\ \citenamefont {Robinson}(2015)}]{linder2015superconducting}%
  \BibitemOpen
  \bibfield  {author} {\bibinfo {author} {\bibfnamefont {J.}~\bibnamefont {Linder}}\ and\ \bibinfo {author} {\bibfnamefont {J.~W.}\ \bibnamefont {Robinson}},\ }\href@noop {} {\bibfield  {journal} {\bibinfo  {journal} {Nature Physics}\ }\textbf {\bibinfo {volume} {11}},\ \bibinfo {pages} {307} (\bibinfo {year} {2015})}\BibitemShut {NoStop}%
\bibitem [{\citenamefont {Maggiora}\ \emph {et~al.}(2024)\citenamefont {Maggiora}, \citenamefont {Wang},\ and\ \citenamefont {Zheng}}]{maggiora2024superconductivity}%
  \BibitemOpen
  \bibfield  {author} {\bibinfo {author} {\bibfnamefont {J.}~\bibnamefont {Maggiora}}, \bibinfo {author} {\bibfnamefont {X.}~\bibnamefont {Wang}}, \ and\ \bibinfo {author} {\bibfnamefont {R.}~\bibnamefont {Zheng}},\ }\href@noop {} {\bibfield  {journal} {\bibinfo  {journal} {Physics Reports}\ }\textbf {\bibinfo {volume} {1076}},\ \bibinfo {pages} {1} (\bibinfo {year} {2024})}\BibitemShut {NoStop}%
\bibitem [{\citenamefont {Yuan}\ \emph {et~al.}(2014)\citenamefont {Yuan}, \citenamefont {Mak},\ and\ \citenamefont {Law}}]{yuan2014possible}%
  \BibitemOpen
  \bibfield  {author} {\bibinfo {author} {\bibfnamefont {N.~F.}\ \bibnamefont {Yuan}}, \bibinfo {author} {\bibfnamefont {K.~F.}\ \bibnamefont {Mak}}, \ and\ \bibinfo {author} {\bibfnamefont {K.~T.}\ \bibnamefont {Law}},\ }\href@noop {} {\bibfield  {journal} {\bibinfo  {journal} {Physical review letters}\ }\textbf {\bibinfo {volume} {113}},\ \bibinfo {pages} {097001} (\bibinfo {year} {2014})}\BibitemShut {NoStop}%
\bibitem [{\citenamefont {Gor'kov}\ and\ \citenamefont {Rashba}(2001)}]{gor2001superconducting}%
  \BibitemOpen
  \bibfield  {author} {\bibinfo {author} {\bibfnamefont {L.~P.}\ \bibnamefont {Gor'kov}}\ and\ \bibinfo {author} {\bibfnamefont {E.~I.}\ \bibnamefont {Rashba}},\ }\href@noop {} {\bibfield  {journal} {\bibinfo  {journal} {Physical Review Letters}\ }\textbf {\bibinfo {volume} {87}},\ \bibinfo {pages} {037004} (\bibinfo {year} {2001})}\BibitemShut {NoStop}%
\bibitem [{\citenamefont {Frigeri}\ \emph {et~al.}(2004{\natexlab{a}})\citenamefont {Frigeri}, \citenamefont {Agterberg}, \citenamefont {Koga},\ and\ \citenamefont {Sigrist}}]{frigeri2004superconductivity}%
  \BibitemOpen
  \bibfield  {author} {\bibinfo {author} {\bibfnamefont {P.}~\bibnamefont {Frigeri}}, \bibinfo {author} {\bibfnamefont {D.}~\bibnamefont {Agterberg}}, \bibinfo {author} {\bibfnamefont {A.}~\bibnamefont {Koga}}, \ and\ \bibinfo {author} {\bibfnamefont {M.}~\bibnamefont {Sigrist}},\ }\href@noop {} {\bibfield  {journal} {\bibinfo  {journal} {Physical review letters}\ }\textbf {\bibinfo {volume} {92}},\ \bibinfo {pages} {097001} (\bibinfo {year} {2004}{\natexlab{a}})}\BibitemShut {NoStop}%
\bibitem [{\citenamefont {Lu}\ \emph {et~al.}(2015)\citenamefont {Lu}, \citenamefont {Zheliuk}, \citenamefont {Leermakers}, \citenamefont {Yuan}, \citenamefont {Zeitler}, \citenamefont {Law},\ and\ \citenamefont {Ye}}]{lu2015evidence}%
  \BibitemOpen
  \bibfield  {author} {\bibinfo {author} {\bibfnamefont {J.}~\bibnamefont {Lu}}, \bibinfo {author} {\bibfnamefont {O.}~\bibnamefont {Zheliuk}}, \bibinfo {author} {\bibfnamefont {I.}~\bibnamefont {Leermakers}}, \bibinfo {author} {\bibfnamefont {N.~F.}\ \bibnamefont {Yuan}}, \bibinfo {author} {\bibfnamefont {U.}~\bibnamefont {Zeitler}}, \bibinfo {author} {\bibfnamefont {K.~T.}\ \bibnamefont {Law}}, \ and\ \bibinfo {author} {\bibfnamefont {J.}~\bibnamefont {Ye}},\ }\href@noop {} {\bibfield  {journal} {\bibinfo  {journal} {Science}\ }\textbf {\bibinfo {volume} {350}},\ \bibinfo {pages} {1353} (\bibinfo {year} {2015})}\BibitemShut {NoStop}%
\bibitem [{\citenamefont {Xi}\ \emph {et~al.}(2016)\citenamefont {Xi}, \citenamefont {Wang}, \citenamefont {Zhao}, \citenamefont {Park}, \citenamefont {Law}, \citenamefont {Berger}, \citenamefont {Forr{\'o}}, \citenamefont {Shan},\ and\ \citenamefont {Mak}}]{xi2016ising}%
  \BibitemOpen
  \bibfield  {author} {\bibinfo {author} {\bibfnamefont {X.}~\bibnamefont {Xi}}, \bibinfo {author} {\bibfnamefont {Z.}~\bibnamefont {Wang}}, \bibinfo {author} {\bibfnamefont {W.}~\bibnamefont {Zhao}}, \bibinfo {author} {\bibfnamefont {J.-H.}\ \bibnamefont {Park}}, \bibinfo {author} {\bibfnamefont {K.~T.}\ \bibnamefont {Law}}, \bibinfo {author} {\bibfnamefont {H.}~\bibnamefont {Berger}}, \bibinfo {author} {\bibfnamefont {L.}~\bibnamefont {Forr{\'o}}}, \bibinfo {author} {\bibfnamefont {J.}~\bibnamefont {Shan}}, \ and\ \bibinfo {author} {\bibfnamefont {K.~F.}\ \bibnamefont {Mak}},\ }\href@noop {} {\bibfield  {journal} {\bibinfo  {journal} {Nature Physics}\ }\textbf {\bibinfo {volume} {12}},\ \bibinfo {pages} {139} (\bibinfo {year} {2016})}\BibitemShut {NoStop}%
\bibitem [{\citenamefont {De~la Barrera}\ \emph {et~al.}(2018)\citenamefont {De~la Barrera}, \citenamefont {Sinko}, \citenamefont {Gopalan}, \citenamefont {Sivadas}, \citenamefont {Seyler}, \citenamefont {Watanabe}, \citenamefont {Taniguchi}, \citenamefont {Tsen}, \citenamefont {Xu}, \citenamefont {Xiao} \emph {et~al.}}]{de2018tuning}%
  \BibitemOpen
  \bibfield  {author} {\bibinfo {author} {\bibfnamefont {S.~C.}\ \bibnamefont {De~la Barrera}}, \bibinfo {author} {\bibfnamefont {M.~R.}\ \bibnamefont {Sinko}}, \bibinfo {author} {\bibfnamefont {D.~P.}\ \bibnamefont {Gopalan}}, \bibinfo {author} {\bibfnamefont {N.}~\bibnamefont {Sivadas}}, \bibinfo {author} {\bibfnamefont {K.~L.}\ \bibnamefont {Seyler}}, \bibinfo {author} {\bibfnamefont {K.}~\bibnamefont {Watanabe}}, \bibinfo {author} {\bibfnamefont {T.}~\bibnamefont {Taniguchi}}, \bibinfo {author} {\bibfnamefont {A.~W.}\ \bibnamefont {Tsen}}, \bibinfo {author} {\bibfnamefont {X.}~\bibnamefont {Xu}}, \bibinfo {author} {\bibfnamefont {D.}~\bibnamefont {Xiao}},  \emph {et~al.},\ }\href@noop {} {\bibfield  {journal} {\bibinfo  {journal} {Nature communications}\ }\textbf {\bibinfo {volume} {9}},\ \bibinfo {pages} {1427} (\bibinfo {year} {2018})}\BibitemShut {NoStop}%
\bibitem [{\citenamefont {Wickramaratne}\ \emph {et~al.}(2020)\citenamefont {Wickramaratne}, \citenamefont {Khmelevskyi}, \citenamefont {Agterberg},\ and\ \citenamefont {Mazin}}]{wickramaratne2020ising}%
  \BibitemOpen
  \bibfield  {author} {\bibinfo {author} {\bibfnamefont {D.}~\bibnamefont {Wickramaratne}}, \bibinfo {author} {\bibfnamefont {S.}~\bibnamefont {Khmelevskyi}}, \bibinfo {author} {\bibfnamefont {D.~F.}\ \bibnamefont {Agterberg}}, \ and\ \bibinfo {author} {\bibfnamefont {I.}~\bibnamefont {Mazin}},\ }\href@noop {} {\bibfield  {journal} {\bibinfo  {journal} {Physical Review X}\ }\textbf {\bibinfo {volume} {10}},\ \bibinfo {pages} {041003} (\bibinfo {year} {2020})}\BibitemShut {NoStop}%
\bibitem [{\citenamefont {Sohn}\ \emph {et~al.}(2018)\citenamefont {Sohn}, \citenamefont {Xi}, \citenamefont {He}, \citenamefont {Jiang}, \citenamefont {Wang}, \citenamefont {Kang}, \citenamefont {Park}, \citenamefont {Berger}, \citenamefont {Forr{\'o}}, \citenamefont {Law} \emph {et~al.}}]{sohn2018unusual}%
  \BibitemOpen
  \bibfield  {author} {\bibinfo {author} {\bibfnamefont {E.}~\bibnamefont {Sohn}}, \bibinfo {author} {\bibfnamefont {X.}~\bibnamefont {Xi}}, \bibinfo {author} {\bibfnamefont {W.-Y.}\ \bibnamefont {He}}, \bibinfo {author} {\bibfnamefont {S.}~\bibnamefont {Jiang}}, \bibinfo {author} {\bibfnamefont {Z.}~\bibnamefont {Wang}}, \bibinfo {author} {\bibfnamefont {K.}~\bibnamefont {Kang}}, \bibinfo {author} {\bibfnamefont {J.-H.}\ \bibnamefont {Park}}, \bibinfo {author} {\bibfnamefont {H.}~\bibnamefont {Berger}}, \bibinfo {author} {\bibfnamefont {L.}~\bibnamefont {Forr{\'o}}}, \bibinfo {author} {\bibfnamefont {K.~T.}\ \bibnamefont {Law}},  \emph {et~al.},\ }\href@noop {} {\bibfield  {journal} {\bibinfo  {journal} {Nature materials}\ }\textbf {\bibinfo {volume} {17}},\ \bibinfo {pages} {504} (\bibinfo {year} {2018})}\BibitemShut {NoStop}%
\bibitem [{\citenamefont {Wang}\ \emph {et~al.}(2019)\citenamefont {Wang}, \citenamefont {Lian}, \citenamefont {Guo}, \citenamefont {Mao}, \citenamefont {Zhang}, \citenamefont {Zhang}, \citenamefont {Gu}, \citenamefont {Xu},\ and\ \citenamefont {Duan}}]{wang2019type}%
  \BibitemOpen
  \bibfield  {author} {\bibinfo {author} {\bibfnamefont {C.}~\bibnamefont {Wang}}, \bibinfo {author} {\bibfnamefont {B.}~\bibnamefont {Lian}}, \bibinfo {author} {\bibfnamefont {X.}~\bibnamefont {Guo}}, \bibinfo {author} {\bibfnamefont {J.}~\bibnamefont {Mao}}, \bibinfo {author} {\bibfnamefont {Z.}~\bibnamefont {Zhang}}, \bibinfo {author} {\bibfnamefont {D.}~\bibnamefont {Zhang}}, \bibinfo {author} {\bibfnamefont {B.-L.}\ \bibnamefont {Gu}}, \bibinfo {author} {\bibfnamefont {Y.}~\bibnamefont {Xu}}, \ and\ \bibinfo {author} {\bibfnamefont {W.}~\bibnamefont {Duan}},\ }\href@noop {} {\bibfield  {journal} {\bibinfo  {journal} {Physical review letters}\ }\textbf {\bibinfo {volume} {123}},\ \bibinfo {pages} {126402} (\bibinfo {year} {2019})}\BibitemShut {NoStop}%
\bibitem [{\citenamefont {Ili{\'c}}\ \emph {et~al.}(2017)\citenamefont {Ili{\'c}}, \citenamefont {Meyer},\ and\ \citenamefont {Houzet}}]{ilic2017enhancement}%
  \BibitemOpen
  \bibfield  {author} {\bibinfo {author} {\bibfnamefont {S.}~\bibnamefont {Ili{\'c}}}, \bibinfo {author} {\bibfnamefont {J.~S.}\ \bibnamefont {Meyer}}, \ and\ \bibinfo {author} {\bibfnamefont {M.}~\bibnamefont {Houzet}},\ }\href@noop {} {\bibfield  {journal} {\bibinfo  {journal} {Physical Review Letters}\ }\textbf {\bibinfo {volume} {119}},\ \bibinfo {pages} {117001} (\bibinfo {year} {2017})}\BibitemShut {NoStop}%
\bibitem [{\citenamefont {Hsu}\ \emph {et~al.}(2017)\citenamefont {Hsu}, \citenamefont {Vaezi}, \citenamefont {Fischer},\ and\ \citenamefont {Kim}}]{hsu2017topological}%
  \BibitemOpen
  \bibfield  {author} {\bibinfo {author} {\bibfnamefont {Y.-T.}\ \bibnamefont {Hsu}}, \bibinfo {author} {\bibfnamefont {A.}~\bibnamefont {Vaezi}}, \bibinfo {author} {\bibfnamefont {M.~H.}\ \bibnamefont {Fischer}}, \ and\ \bibinfo {author} {\bibfnamefont {E.-A.}\ \bibnamefont {Kim}},\ }\href@noop {} {\bibfield  {journal} {\bibinfo  {journal} {Nature communications}\ }\textbf {\bibinfo {volume} {8}},\ \bibinfo {pages} {1} (\bibinfo {year} {2017})}\BibitemShut {NoStop}%
\bibitem [{\citenamefont {Nakamura}\ and\ \citenamefont {Yanase}(2017)}]{nakamura2017odd}%
  \BibitemOpen
  \bibfield  {author} {\bibinfo {author} {\bibfnamefont {Y.}~\bibnamefont {Nakamura}}\ and\ \bibinfo {author} {\bibfnamefont {Y.}~\bibnamefont {Yanase}},\ }\href@noop {} {\bibfield  {journal} {\bibinfo  {journal} {Physical Review B}\ }\textbf {\bibinfo {volume} {96}},\ \bibinfo {pages} {054501} (\bibinfo {year} {2017})}\BibitemShut {NoStop}%
\bibitem [{\citenamefont {Zhou}\ \emph {et~al.}(2016)\citenamefont {Zhou}, \citenamefont {Yuan},\ and\ \citenamefont {Jiang}}]{zhou2016ising}%
  \BibitemOpen
  \bibfield  {author} {\bibinfo {author} {\bibfnamefont {B.~T.}\ \bibnamefont {Zhou}}, \bibinfo {author} {\bibfnamefont {N.~F.}\ \bibnamefont {Yuan}}, \ and\ \bibinfo {author} {\bibfnamefont {H.-L.}\ \bibnamefont {Jiang}},\ }\href@noop {} {\bibfield  {journal} {\bibinfo  {journal} {Physical Review B}\ }\textbf {\bibinfo {volume} {93}},\ \bibinfo {pages} {180501} (\bibinfo {year} {2016})}\BibitemShut {NoStop}%
\bibitem [{\citenamefont {He}\ \emph {et~al.}(2018)\citenamefont {He}, \citenamefont {Zhou}, \citenamefont {He}, \citenamefont {Yuan}, \citenamefont {Zhang},\ and\ \citenamefont {Law}}]{he2018magnetic}%
  \BibitemOpen
  \bibfield  {author} {\bibinfo {author} {\bibfnamefont {W.-Y.}\ \bibnamefont {He}}, \bibinfo {author} {\bibfnamefont {B.~T.}\ \bibnamefont {Zhou}}, \bibinfo {author} {\bibfnamefont {J.~J.}\ \bibnamefont {He}}, \bibinfo {author} {\bibfnamefont {N.~F.}\ \bibnamefont {Yuan}}, \bibinfo {author} {\bibfnamefont {T.}~\bibnamefont {Zhang}}, \ and\ \bibinfo {author} {\bibfnamefont {K.~T.}\ \bibnamefont {Law}},\ }\href@noop {} {\bibfield  {journal} {\bibinfo  {journal} {Communications Physics}\ }\textbf {\bibinfo {volume} {1}},\ \bibinfo {pages} {40} (\bibinfo {year} {2018})}\BibitemShut {NoStop}%
\bibitem [{\citenamefont {Xie}\ \emph {et~al.}(2020{\natexlab{a}})\citenamefont {Xie}, \citenamefont {Zhou}, \citenamefont {Ng},\ and\ \citenamefont {Law}}]{xie2020strongly}%
  \BibitemOpen
  \bibfield  {author} {\bibinfo {author} {\bibfnamefont {Y.}~\bibnamefont {Xie}}, \bibinfo {author} {\bibfnamefont {B.~T.}\ \bibnamefont {Zhou}}, \bibinfo {author} {\bibfnamefont {T.~K.}\ \bibnamefont {Ng}}, \ and\ \bibinfo {author} {\bibfnamefont {K.~T.}\ \bibnamefont {Law}},\ }\href@noop {} {\bibfield  {journal} {\bibinfo  {journal} {Physical Review Research}\ }\textbf {\bibinfo {volume} {2}},\ \bibinfo {pages} {013026} (\bibinfo {year} {2020}{\natexlab{a}})}\BibitemShut {NoStop}%
\bibitem [{\citenamefont {Xie}\ and\ \citenamefont {Law}(2023)}]{xie2023orbital}%
  \BibitemOpen
  \bibfield  {author} {\bibinfo {author} {\bibfnamefont {Y.-M.}\ \bibnamefont {Xie}}\ and\ \bibinfo {author} {\bibfnamefont {K.~T.}\ \bibnamefont {Law}},\ }\href@noop {} {\bibfield  {journal} {\bibinfo  {journal} {Physical Review Letters}\ }\textbf {\bibinfo {volume} {131}},\ \bibinfo {pages} {016001} (\bibinfo {year} {2023})}\BibitemShut {NoStop}%
\bibitem [{\citenamefont {{\v{S}}mejkal}\ \emph {et~al.}(2022{\natexlab{a}})\citenamefont {{\v{S}}mejkal}, \citenamefont {Sinova},\ and\ \citenamefont {Jungwirth}}]{vsmejkal2022emerging}%
  \BibitemOpen
  \bibfield  {author} {\bibinfo {author} {\bibfnamefont {L.}~\bibnamefont {{\v{S}}mejkal}}, \bibinfo {author} {\bibfnamefont {J.}~\bibnamefont {Sinova}}, \ and\ \bibinfo {author} {\bibfnamefont {T.}~\bibnamefont {Jungwirth}},\ }\href@noop {} {\bibfield  {journal} {\bibinfo  {journal} {Physical Review X}\ }\textbf {\bibinfo {volume} {12}},\ \bibinfo {pages} {040501} (\bibinfo {year} {2022}{\natexlab{a}})}\BibitemShut {NoStop}%
\bibitem [{\citenamefont {Ma}\ \emph {et~al.}(2021)\citenamefont {Ma}, \citenamefont {Hu}, \citenamefont {Li}, \citenamefont {Liu}, \citenamefont {Yao}, \citenamefont {Jia},\ and\ \citenamefont {Liu}}]{ma2021multifunctional}%
  \BibitemOpen
  \bibfield  {author} {\bibinfo {author} {\bibfnamefont {H.-Y.}\ \bibnamefont {Ma}}, \bibinfo {author} {\bibfnamefont {M.}~\bibnamefont {Hu}}, \bibinfo {author} {\bibfnamefont {N.}~\bibnamefont {Li}}, \bibinfo {author} {\bibfnamefont {J.}~\bibnamefont {Liu}}, \bibinfo {author} {\bibfnamefont {W.}~\bibnamefont {Yao}}, \bibinfo {author} {\bibfnamefont {J.-F.}\ \bibnamefont {Jia}}, \ and\ \bibinfo {author} {\bibfnamefont {J.}~\bibnamefont {Liu}},\ }\href@noop {} {\bibfield  {journal} {\bibinfo  {journal} {Nature communications}\ }\textbf {\bibinfo {volume} {12}},\ \bibinfo {pages} {2846} (\bibinfo {year} {2021})}\BibitemShut {NoStop}%
\bibitem [{\citenamefont {Lee}\ \emph {et~al.}(2024{\natexlab{a}})\citenamefont {Lee}, \citenamefont {Lee}, \citenamefont {Jung}, \citenamefont {Jung}, \citenamefont {Kim}, \citenamefont {Lee}, \citenamefont {Seok}, \citenamefont {Kim}, \citenamefont {Park}, \citenamefont {{\v{S}}mejkal} \emph {et~al.}}]{lee2024broken}%
  \BibitemOpen
  \bibfield  {author} {\bibinfo {author} {\bibfnamefont {S.}~\bibnamefont {Lee}}, \bibinfo {author} {\bibfnamefont {S.}~\bibnamefont {Lee}}, \bibinfo {author} {\bibfnamefont {S.}~\bibnamefont {Jung}}, \bibinfo {author} {\bibfnamefont {J.}~\bibnamefont {Jung}}, \bibinfo {author} {\bibfnamefont {D.}~\bibnamefont {Kim}}, \bibinfo {author} {\bibfnamefont {Y.}~\bibnamefont {Lee}}, \bibinfo {author} {\bibfnamefont {B.}~\bibnamefont {Seok}}, \bibinfo {author} {\bibfnamefont {J.}~\bibnamefont {Kim}}, \bibinfo {author} {\bibfnamefont {B.~G.}\ \bibnamefont {Park}}, \bibinfo {author} {\bibfnamefont {L.}~\bibnamefont {{\v{S}}mejkal}},  \emph {et~al.},\ }\href@noop {} {\bibfield  {journal} {\bibinfo  {journal} {Physical Review Letters}\ }\textbf {\bibinfo {volume} {132}},\ \bibinfo {pages} {036702} (\bibinfo {year} {2024}{\natexlab{a}})}\BibitemShut {NoStop}%
\bibitem [{\citenamefont {Mazin}\ \emph {et~al.}(2022)\citenamefont {Mazin} \emph {et~al.}}]{mazin2022altermagnetism}%
  \BibitemOpen
  \bibfield  {author} {\bibinfo {author} {\bibfnamefont {I.}~\bibnamefont {Mazin}} \emph {et~al.},\ }\href@noop {} {\bibfield  {journal} {\bibinfo  {journal} {Physical Review X}\ }\textbf {\bibinfo {volume} {12}},\ \bibinfo {pages} {040002} (\bibinfo {year} {2022})}\BibitemShut {NoStop}%
\bibitem [{\citenamefont {{\v{S}}mejkal}\ \emph {et~al.}(2022{\natexlab{b}})\citenamefont {{\v{S}}mejkal}, \citenamefont {Hellenes}, \citenamefont {Gonz{\'a}lez-Hern{\'a}ndez}, \citenamefont {Sinova},\ and\ \citenamefont {Jungwirth}}]{vsmejkal2022giant}%
  \BibitemOpen
  \bibfield  {author} {\bibinfo {author} {\bibfnamefont {L.}~\bibnamefont {{\v{S}}mejkal}}, \bibinfo {author} {\bibfnamefont {A.~B.}\ \bibnamefont {Hellenes}}, \bibinfo {author} {\bibfnamefont {R.}~\bibnamefont {Gonz{\'a}lez-Hern{\'a}ndez}}, \bibinfo {author} {\bibfnamefont {J.}~\bibnamefont {Sinova}}, \ and\ \bibinfo {author} {\bibfnamefont {T.}~\bibnamefont {Jungwirth}},\ }\href@noop {} {\bibfield  {journal} {\bibinfo  {journal} {Physical Review X}\ }\textbf {\bibinfo {volume} {12}},\ \bibinfo {pages} {011028} (\bibinfo {year} {2022}{\natexlab{b}})}\BibitemShut {NoStop}%
\bibitem [{\citenamefont {Ghorashi}\ \emph {et~al.}(2024)\citenamefont {Ghorashi}, \citenamefont {Hughes},\ and\ \citenamefont {Cano}}]{ghorashi2024altermagnetic}%
  \BibitemOpen
  \bibfield  {author} {\bibinfo {author} {\bibfnamefont {S.~A.~A.}\ \bibnamefont {Ghorashi}}, \bibinfo {author} {\bibfnamefont {T.~L.}\ \bibnamefont {Hughes}}, \ and\ \bibinfo {author} {\bibfnamefont {J.}~\bibnamefont {Cano}},\ }\href@noop {} {\bibfield  {journal} {\bibinfo  {journal} {Physical Review Letters}\ }\textbf {\bibinfo {volume} {133}},\ \bibinfo {pages} {106601} (\bibinfo {year} {2024})}\BibitemShut {NoStop}%
\bibitem [{\citenamefont {Fang}\ \emph {et~al.}(2024)\citenamefont {Fang}, \citenamefont {Cano},\ and\ \citenamefont {Ghorashi}}]{fang2024quantum}%
  \BibitemOpen
  \bibfield  {author} {\bibinfo {author} {\bibfnamefont {Y.}~\bibnamefont {Fang}}, \bibinfo {author} {\bibfnamefont {J.}~\bibnamefont {Cano}}, \ and\ \bibinfo {author} {\bibfnamefont {S.~A.~A.}\ \bibnamefont {Ghorashi}},\ }\href@noop {} {\bibfield  {journal} {\bibinfo  {journal} {Physical Review Letters}\ }\textbf {\bibinfo {volume} {133}},\ \bibinfo {pages} {106701} (\bibinfo {year} {2024})}\BibitemShut {NoStop}%
\bibitem [{\citenamefont {Xiao}\ \emph {et~al.}(2024)\citenamefont {Xiao}, \citenamefont {Zhao}, \citenamefont {Li}, \citenamefont {Shindou},\ and\ \citenamefont {Song}}]{xiao2024spin}%
  \BibitemOpen
  \bibfield  {author} {\bibinfo {author} {\bibfnamefont {Z.}~\bibnamefont {Xiao}}, \bibinfo {author} {\bibfnamefont {J.}~\bibnamefont {Zhao}}, \bibinfo {author} {\bibfnamefont {Y.}~\bibnamefont {Li}}, \bibinfo {author} {\bibfnamefont {R.}~\bibnamefont {Shindou}}, \ and\ \bibinfo {author} {\bibfnamefont {Z.-D.}\ \bibnamefont {Song}},\ }\href@noop {} {\bibfield  {journal} {\bibinfo  {journal} {Physical Review X}\ }\textbf {\bibinfo {volume} {14}},\ \bibinfo {pages} {031037} (\bibinfo {year} {2024})}\BibitemShut {NoStop}%
\bibitem [{\citenamefont {Jiang}\ \emph {et~al.}(2024)\citenamefont {Jiang}, \citenamefont {Song}, \citenamefont {Zhu}, \citenamefont {Fang}, \citenamefont {Weng}, \citenamefont {Liu}, \citenamefont {Yang},\ and\ \citenamefont {Fang}}]{jiang2024enumeration}%
  \BibitemOpen
  \bibfield  {author} {\bibinfo {author} {\bibfnamefont {Y.}~\bibnamefont {Jiang}}, \bibinfo {author} {\bibfnamefont {Z.}~\bibnamefont {Song}}, \bibinfo {author} {\bibfnamefont {T.}~\bibnamefont {Zhu}}, \bibinfo {author} {\bibfnamefont {Z.}~\bibnamefont {Fang}}, \bibinfo {author} {\bibfnamefont {H.}~\bibnamefont {Weng}}, \bibinfo {author} {\bibfnamefont {Z.-X.}\ \bibnamefont {Liu}}, \bibinfo {author} {\bibfnamefont {J.}~\bibnamefont {Yang}}, \ and\ \bibinfo {author} {\bibfnamefont {C.}~\bibnamefont {Fang}},\ }\href@noop {} {\bibfield  {journal} {\bibinfo  {journal} {Physical Review X}\ }\textbf {\bibinfo {volume} {14}},\ \bibinfo {pages} {031039} (\bibinfo {year} {2024})}\BibitemShut {NoStop}%
\bibitem [{\citenamefont {Chen}\ \emph {et~al.}(2024)\citenamefont {Chen}, \citenamefont {Ren}, \citenamefont {Zhu}, \citenamefont {Yu}, \citenamefont {Zhang}, \citenamefont {Liu}, \citenamefont {Li}, \citenamefont {Liu}, \citenamefont {Li},\ and\ \citenamefont {Liu}}]{chen2024enumeration}%
  \BibitemOpen
  \bibfield  {author} {\bibinfo {author} {\bibfnamefont {X.}~\bibnamefont {Chen}}, \bibinfo {author} {\bibfnamefont {J.}~\bibnamefont {Ren}}, \bibinfo {author} {\bibfnamefont {Y.}~\bibnamefont {Zhu}}, \bibinfo {author} {\bibfnamefont {Y.}~\bibnamefont {Yu}}, \bibinfo {author} {\bibfnamefont {A.}~\bibnamefont {Zhang}}, \bibinfo {author} {\bibfnamefont {P.}~\bibnamefont {Liu}}, \bibinfo {author} {\bibfnamefont {J.}~\bibnamefont {Li}}, \bibinfo {author} {\bibfnamefont {Y.}~\bibnamefont {Liu}}, \bibinfo {author} {\bibfnamefont {C.}~\bibnamefont {Li}}, \ and\ \bibinfo {author} {\bibfnamefont {Q.}~\bibnamefont {Liu}},\ }\href@noop {} {\bibfield  {journal} {\bibinfo  {journal} {Physical Review X}\ }\textbf {\bibinfo {volume} {14}},\ \bibinfo {pages} {031038} (\bibinfo {year} {2024})}\BibitemShut {NoStop}%
\bibitem [{\citenamefont {Lee}\ \emph {et~al.}(2024{\natexlab{b}})\citenamefont {Lee}, \citenamefont {Qian},\ and\ \citenamefont {Yang}}]{lee2024fermi}%
  \BibitemOpen
  \bibfield  {author} {\bibinfo {author} {\bibfnamefont {S.~H.}\ \bibnamefont {Lee}}, \bibinfo {author} {\bibfnamefont {Y.}~\bibnamefont {Qian}}, \ and\ \bibinfo {author} {\bibfnamefont {B.-J.}\ \bibnamefont {Yang}},\ }\href@noop {} {\bibfield  {journal} {\bibinfo  {journal} {Physical Review Letters}\ }\textbf {\bibinfo {volume} {132}},\ \bibinfo {pages} {196602} (\bibinfo {year} {2024}{\natexlab{b}})}\BibitemShut {NoStop}%
\bibitem [{\citenamefont {Papaj}(2023)}]{papaj2023andreev}%
  \BibitemOpen
  \bibfield  {author} {\bibinfo {author} {\bibfnamefont {M.}~\bibnamefont {Papaj}},\ }\href@noop {} {\bibfield  {journal} {\bibinfo  {journal} {Physical Review B}\ }\textbf {\bibinfo {volume} {108}},\ \bibinfo {pages} {L060508} (\bibinfo {year} {2023})}\BibitemShut {NoStop}%
\bibitem [{\citenamefont {Sun}\ \emph {et~al.}(2023)\citenamefont {Sun}, \citenamefont {Brataas},\ and\ \citenamefont {Linder}}]{sun2023andreev}%
  \BibitemOpen
  \bibfield  {author} {\bibinfo {author} {\bibfnamefont {C.}~\bibnamefont {Sun}}, \bibinfo {author} {\bibfnamefont {A.}~\bibnamefont {Brataas}}, \ and\ \bibinfo {author} {\bibfnamefont {J.}~\bibnamefont {Linder}},\ }\href@noop {} {\bibfield  {journal} {\bibinfo  {journal} {Physical Review B}\ }\textbf {\bibinfo {volume} {108}},\ \bibinfo {pages} {054511} (\bibinfo {year} {2023})}\BibitemShut {NoStop}%
\bibitem [{\citenamefont {Ouassou}\ \emph {et~al.}(2023)\citenamefont {Ouassou}, \citenamefont {Brataas},\ and\ \citenamefont {Linder}}]{ouassou2023dc}%
  \BibitemOpen
  \bibfield  {author} {\bibinfo {author} {\bibfnamefont {J.~A.}\ \bibnamefont {Ouassou}}, \bibinfo {author} {\bibfnamefont {A.}~\bibnamefont {Brataas}}, \ and\ \bibinfo {author} {\bibfnamefont {J.}~\bibnamefont {Linder}},\ }\href@noop {} {\bibfield  {journal} {\bibinfo  {journal} {Physical Review Letters}\ }\textbf {\bibinfo {volume} {131}},\ \bibinfo {pages} {076003} (\bibinfo {year} {2023})}\BibitemShut {NoStop}%
\bibitem [{\citenamefont {Beenakker}\ and\ \citenamefont {Vakhtel}(2023)}]{beenakker2023phase}%
  \BibitemOpen
  \bibfield  {author} {\bibinfo {author} {\bibfnamefont {C.}~\bibnamefont {Beenakker}}\ and\ \bibinfo {author} {\bibfnamefont {T.}~\bibnamefont {Vakhtel}},\ }\href@noop {} {\bibfield  {journal} {\bibinfo  {journal} {Physical Review B}\ }\textbf {\bibinfo {volume} {108}},\ \bibinfo {pages} {075425} (\bibinfo {year} {2023})}\BibitemShut {NoStop}%
\bibitem [{\citenamefont {Lu}\ \emph {et~al.}(2024)\citenamefont {Lu}, \citenamefont {Maeda}, \citenamefont {Ito}, \citenamefont {Yada},\ and\ \citenamefont {Tanaka}}]{lu2024varphi}%
  \BibitemOpen
  \bibfield  {author} {\bibinfo {author} {\bibfnamefont {B.}~\bibnamefont {Lu}}, \bibinfo {author} {\bibfnamefont {K.}~\bibnamefont {Maeda}}, \bibinfo {author} {\bibfnamefont {H.}~\bibnamefont {Ito}}, \bibinfo {author} {\bibfnamefont {K.}~\bibnamefont {Yada}}, \ and\ \bibinfo {author} {\bibfnamefont {Y.}~\bibnamefont {Tanaka}},\ }\href@noop {} {\bibfield  {journal} {\bibinfo  {journal} {Physical Review Letters}\ }\textbf {\bibinfo {volume} {133}},\ \bibinfo {pages} {226002} (\bibinfo {year} {2024})}\BibitemShut {NoStop}%
\bibitem [{\citenamefont {Yang}\ \emph {et~al.}(2025)\citenamefont {Yang}, \citenamefont {Sun}, \citenamefont {Xie},\ and\ \citenamefont {Law}}]{yang2025topological}%
  \BibitemOpen
  \bibfield  {author} {\bibinfo {author} {\bibfnamefont {G.~Z.~X.}\ \bibnamefont {Yang}}, \bibinfo {author} {\bibfnamefont {Z.-T.}\ \bibnamefont {Sun}}, \bibinfo {author} {\bibfnamefont {Y.-M.}\ \bibnamefont {Xie}}, \ and\ \bibinfo {author} {\bibfnamefont {K.~T.}\ \bibnamefont {Law}},\ }\href@noop {} {\bibfield  {journal} {\bibinfo  {journal} {arXiv preprint arXiv:2502.20283}\ } (\bibinfo {year} {2025})}\BibitemShut {NoStop}%
\bibitem [{\citenamefont {Zhang}\ \emph {et~al.}(2024)\citenamefont {Zhang}, \citenamefont {Hu},\ and\ \citenamefont {Neupert}}]{zhang2024finite}%
  \BibitemOpen
  \bibfield  {author} {\bibinfo {author} {\bibfnamefont {S.-B.}\ \bibnamefont {Zhang}}, \bibinfo {author} {\bibfnamefont {L.-H.}\ \bibnamefont {Hu}}, \ and\ \bibinfo {author} {\bibfnamefont {T.}~\bibnamefont {Neupert}},\ }\href@noop {} {\bibfield  {journal} {\bibinfo  {journal} {Nature Communications}\ }\textbf {\bibinfo {volume} {15}},\ \bibinfo {pages} {1801} (\bibinfo {year} {2024})}\BibitemShut {NoStop}%
\bibitem [{\citenamefont {Hu}\ \emph {et~al.}(2025)\citenamefont {Hu}, \citenamefont {Matsyshyn},\ and\ \citenamefont {Song}}]{hu2025nonlinear}%
  \BibitemOpen
  \bibfield  {author} {\bibinfo {author} {\bibfnamefont {J.-X.}\ \bibnamefont {Hu}}, \bibinfo {author} {\bibfnamefont {O.}~\bibnamefont {Matsyshyn}}, \ and\ \bibinfo {author} {\bibfnamefont {J.~C.}\ \bibnamefont {Song}},\ }\href@noop {} {\bibfield  {journal} {\bibinfo  {journal} {Physical Review Letters}\ }\textbf {\bibinfo {volume} {134}},\ \bibinfo {pages} {026001} (\bibinfo {year} {2025})}\BibitemShut {NoStop}%
\bibitem [{\citenamefont {Zyuzin}(2024)}]{zyuzin2024magnetoelectric}%
  \BibitemOpen
  \bibfield  {author} {\bibinfo {author} {\bibfnamefont {A.~A.}\ \bibnamefont {Zyuzin}},\ }\href@noop {} {\bibfield  {journal} {\bibinfo  {journal} {Physical Review B}\ }\textbf {\bibinfo {volume} {109}},\ \bibinfo {pages} {L220505} (\bibinfo {year} {2024})}\BibitemShut {NoStop}%
\bibitem [{\citenamefont {Hellenes}\ \emph {et~al.}(2023)\citenamefont {Hellenes}, \citenamefont {Jungwirth}, \citenamefont {Sinova},\ and\ \citenamefont {{\v{S}}mejkal}}]{hellenes2023exchange}%
  \BibitemOpen
  \bibfield  {author} {\bibinfo {author} {\bibfnamefont {A.~B.}\ \bibnamefont {Hellenes}}, \bibinfo {author} {\bibfnamefont {T.}~\bibnamefont {Jungwirth}}, \bibinfo {author} {\bibfnamefont {J.}~\bibnamefont {Sinova}}, \ and\ \bibinfo {author} {\bibfnamefont {L.}~\bibnamefont {{\v{S}}mejkal}},\ }\href@noop {} {\bibfield  {journal} {\bibinfo  {journal} {arXiv preprint arXiv:2309.01607}\ } (\bibinfo {year} {2023})}\BibitemShut {NoStop}%
\bibitem [{\citenamefont {Brekke}\ \emph {et~al.}(2024)\citenamefont {Brekke}, \citenamefont {Sukhachov}, \citenamefont {Giil}, \citenamefont {Brataas},\ and\ \citenamefont {Linder}}]{brekke2024minimal}%
  \BibitemOpen
  \bibfield  {author} {\bibinfo {author} {\bibfnamefont {B.}~\bibnamefont {Brekke}}, \bibinfo {author} {\bibfnamefont {P.}~\bibnamefont {Sukhachov}}, \bibinfo {author} {\bibfnamefont {H.~G.}\ \bibnamefont {Giil}}, \bibinfo {author} {\bibfnamefont {A.}~\bibnamefont {Brataas}}, \ and\ \bibinfo {author} {\bibfnamefont {J.}~\bibnamefont {Linder}},\ }\href@noop {} {\bibfield  {journal} {\bibinfo  {journal} {Physical Review Letters}\ }\textbf {\bibinfo {volume} {133}},\ \bibinfo {pages} {236703} (\bibinfo {year} {2024})}\BibitemShut {NoStop}%
\bibitem [{\citenamefont {Maeda}\ \emph {et~al.}(2024)\citenamefont {Maeda}, \citenamefont {Lu}, \citenamefont {Yada},\ and\ \citenamefont {Tanaka}}]{maeda2024theory}%
  \BibitemOpen
  \bibfield  {author} {\bibinfo {author} {\bibfnamefont {K.}~\bibnamefont {Maeda}}, \bibinfo {author} {\bibfnamefont {B.}~\bibnamefont {Lu}}, \bibinfo {author} {\bibfnamefont {K.}~\bibnamefont {Yada}}, \ and\ \bibinfo {author} {\bibfnamefont {Y.}~\bibnamefont {Tanaka}},\ }\href@noop {} {\bibfield  {journal} {\bibinfo  {journal} {Journal of the Physical Society of Japan}\ }\textbf {\bibinfo {volume} {93}},\ \bibinfo {pages} {114703} (\bibinfo {year} {2024})}\BibitemShut {NoStop}%
\bibitem [{\citenamefont {Sukhachov}\ and\ \citenamefont {Linder}(2024)}]{sukhachov2024impurity}%
  \BibitemOpen
  \bibfield  {author} {\bibinfo {author} {\bibfnamefont {P.}~\bibnamefont {Sukhachov}}\ and\ \bibinfo {author} {\bibfnamefont {J.}~\bibnamefont {Linder}},\ }\href@noop {} {\bibfield  {journal} {\bibinfo  {journal} {Physical Review B}\ }\textbf {\bibinfo {volume} {110}},\ \bibinfo {pages} {205114} (\bibinfo {year} {2024})}\BibitemShut {NoStop}%
\bibitem [{\citenamefont {Chakraborty}\ \emph {et~al.}(2024)\citenamefont {Chakraborty}, \citenamefont {Hellenes}, \citenamefont {Jaeschke-Ubiergo}, \citenamefont {Jungwirth}, \citenamefont {{\v{S}}mejkal},\ and\ \citenamefont {Sinova}}]{chakraborty2024highly}%
  \BibitemOpen
  \bibfield  {author} {\bibinfo {author} {\bibfnamefont {A.}~\bibnamefont {Chakraborty}}, \bibinfo {author} {\bibfnamefont {A.~B.}\ \bibnamefont {Hellenes}}, \bibinfo {author} {\bibfnamefont {R.}~\bibnamefont {Jaeschke-Ubiergo}}, \bibinfo {author} {\bibfnamefont {T.}~\bibnamefont {Jungwirth}}, \bibinfo {author} {\bibfnamefont {L.}~\bibnamefont {{\v{S}}mejkal}}, \ and\ \bibinfo {author} {\bibfnamefont {J.}~\bibnamefont {Sinova}},\ }\href@noop {} {\bibfield  {journal} {\bibinfo  {journal} {arXiv preprint arXiv:2411.16378}\ } (\bibinfo {year} {2024})}\BibitemShut {NoStop}%
\bibitem [{\citenamefont {Kokkeler}\ \emph {et~al.}(2024)\citenamefont {Kokkeler}, \citenamefont {Tokatly},\ and\ \citenamefont {Bergeret}}]{kokkeler2024quantum}%
  \BibitemOpen
  \bibfield  {author} {\bibinfo {author} {\bibfnamefont {T.}~\bibnamefont {Kokkeler}}, \bibinfo {author} {\bibfnamefont {I.}~\bibnamefont {Tokatly}}, \ and\ \bibinfo {author} {\bibfnamefont {F.~S.}\ \bibnamefont {Bergeret}},\ }\href@noop {} {\bibfield  {journal} {\bibinfo  {journal} {arXiv preprint arXiv:2412.10236}\ } (\bibinfo {year} {2024})}\BibitemShut {NoStop}%
\bibitem [{\citenamefont {Sukhachov}\ \emph {et~al.}(2024)\citenamefont {Sukhachov}, \citenamefont {Gl{\o}ckner~Giil}, \citenamefont {Brekke},\ and\ \citenamefont {Linder}}]{sukhachov2024coexistence}%
  \BibitemOpen
  \bibfield  {author} {\bibinfo {author} {\bibfnamefont {P.}~\bibnamefont {Sukhachov}}, \bibinfo {author} {\bibfnamefont {H.}~\bibnamefont {Gl{\o}ckner~Giil}}, \bibinfo {author} {\bibfnamefont {B.}~\bibnamefont {Brekke}}, \ and\ \bibinfo {author} {\bibfnamefont {J.}~\bibnamefont {Linder}},\ }\href@noop {} {\bibfield  {journal} {\bibinfo  {journal} {arXiv preprint arXiv:2412.14245}\ } (\bibinfo {year} {2024})}\BibitemShut {NoStop}%
\bibitem [{\citenamefont {Bergeret}\ \emph {et~al.}(2004)\citenamefont {Bergeret}, \citenamefont {Volkov},\ and\ \citenamefont {Efetov}}]{bergeret2004induced}%
  \BibitemOpen
  \bibfield  {author} {\bibinfo {author} {\bibfnamefont {F.}~\bibnamefont {Bergeret}}, \bibinfo {author} {\bibfnamefont {A.~F.}\ \bibnamefont {Volkov}}, \ and\ \bibinfo {author} {\bibfnamefont {K.}~\bibnamefont {Efetov}},\ }\href@noop {} {\bibfield  {journal} {\bibinfo  {journal} {Physical Review B}\ }\textbf {\bibinfo {volume} {69}},\ \bibinfo {pages} {174504} (\bibinfo {year} {2004})}\BibitemShut {NoStop}%
\bibitem [{Not()}]{NoteX}%
  \BibitemOpen
  \href@noop {} {}\bibinfo {note} {The Supplemental Material of ``Pseudo-Ising superconductivity induced by $p$-wave magnetism", which includes four parts: I. Derivation of the low-energy effective Hamiltonian of $p$-wave magnet; II. Formalism of Edelstein effect; III. Pseudo-Ising superconductivity with $f$-wave magnetism; IV. Calculations for four-band models.}\BibitemShut {Stop}%
\bibitem [{\citenamefont {Frigeri}\ \emph {et~al.}(2004{\natexlab{b}})\citenamefont {Frigeri}, \citenamefont {Agterberg},\ and\ \citenamefont {Sigrist}}]{frigeri2004spin}%
  \BibitemOpen
  \bibfield  {author} {\bibinfo {author} {\bibfnamefont {P.~A.}\ \bibnamefont {Frigeri}}, \bibinfo {author} {\bibfnamefont {D.~F.}\ \bibnamefont {Agterberg}}, \ and\ \bibinfo {author} {\bibfnamefont {M.}~\bibnamefont {Sigrist}},\ }\href@noop {} {\bibfield  {journal} {\bibinfo  {journal} {New Journal of Physics}\ }\textbf {\bibinfo {volume} {6}},\ \bibinfo {pages} {115} (\bibinfo {year} {2004}{\natexlab{b}})}\BibitemShut {NoStop}%
\bibitem [{\citenamefont {Xie}\ \emph {et~al.}(2020{\natexlab{b}})\citenamefont {Xie}, \citenamefont {Zhou},\ and\ \citenamefont {Law}}]{xie2020spin}%
  \BibitemOpen
  \bibfield  {author} {\bibinfo {author} {\bibfnamefont {Y.-M.}\ \bibnamefont {Xie}}, \bibinfo {author} {\bibfnamefont {B.~T.}\ \bibnamefont {Zhou}}, \ and\ \bibinfo {author} {\bibfnamefont {K.~T.}\ \bibnamefont {Law}},\ }\href@noop {} {\bibfield  {journal} {\bibinfo  {journal} {Physical Review Letters}\ }\textbf {\bibinfo {volume} {125}},\ \bibinfo {pages} {107001} (\bibinfo {year} {2020}{\natexlab{b}})}\BibitemShut {NoStop}%
\bibitem [{\citenamefont {Maki}\ and\ \citenamefont {Tsuneto}(1964)}]{maki1964pauli}%
  \BibitemOpen
  \bibfield  {author} {\bibinfo {author} {\bibfnamefont {K.}~\bibnamefont {Maki}}\ and\ \bibinfo {author} {\bibfnamefont {T.}~\bibnamefont {Tsuneto}},\ }\href@noop {} {\bibfield  {journal} {\bibinfo  {journal} {Progress of Theoretical Physics}\ }\textbf {\bibinfo {volume} {31}},\ \bibinfo {pages} {945} (\bibinfo {year} {1964})}\BibitemShut {NoStop}%
\bibitem [{\citenamefont {Edelstein}(1995)}]{edelstein1995magnetoelectric}%
  \BibitemOpen
  \bibfield  {author} {\bibinfo {author} {\bibfnamefont {V.~M.}\ \bibnamefont {Edelstein}},\ }\href@noop {} {\bibfield  {journal} {\bibinfo  {journal} {Physical review letters}\ }\textbf {\bibinfo {volume} {75}},\ \bibinfo {pages} {2004} (\bibinfo {year} {1995})}\BibitemShut {NoStop}%
\bibitem [{\citenamefont {Edelstein}(2005)}]{edelstein2005magnetoelectric}%
  \BibitemOpen
  \bibfield  {author} {\bibinfo {author} {\bibfnamefont {V.~M.}\ \bibnamefont {Edelstein}},\ }\href@noop {} {\bibfield  {journal} {\bibinfo  {journal} {Physical Review B}\ }\textbf {\bibinfo {volume} {72}},\ \bibinfo {pages} {172501} (\bibinfo {year} {2005})}\BibitemShut {NoStop}%
\bibitem [{\citenamefont {Chirolli}\ \emph {et~al.}(2022)\citenamefont {Chirolli}, \citenamefont {Mercaldo}, \citenamefont {Guarcello}, \citenamefont {Giazotto},\ and\ \citenamefont {Cuoco}}]{chirolli2022colossal}%
  \BibitemOpen
  \bibfield  {author} {\bibinfo {author} {\bibfnamefont {L.}~\bibnamefont {Chirolli}}, \bibinfo {author} {\bibfnamefont {M.~T.}\ \bibnamefont {Mercaldo}}, \bibinfo {author} {\bibfnamefont {C.}~\bibnamefont {Guarcello}}, \bibinfo {author} {\bibfnamefont {F.}~\bibnamefont {Giazotto}}, \ and\ \bibinfo {author} {\bibfnamefont {M.}~\bibnamefont {Cuoco}},\ }\href@noop {} {\bibfield  {journal} {\bibinfo  {journal} {Physical Review Letters}\ }\textbf {\bibinfo {volume} {128}},\ \bibinfo {pages} {217703} (\bibinfo {year} {2022})}\BibitemShut {NoStop}%
\bibitem [{\citenamefont {Alicea}(2012)}]{alicea2012new}%
  \BibitemOpen
  \bibfield  {author} {\bibinfo {author} {\bibfnamefont {J.}~\bibnamefont {Alicea}},\ }\href@noop {} {\bibfield  {journal} {\bibinfo  {journal} {Reports on progress in physics}\ }\textbf {\bibinfo {volume} {75}},\ \bibinfo {pages} {076501} (\bibinfo {year} {2012})}\BibitemShut {NoStop}%
\bibitem [{\citenamefont {Beenakker}(2013)}]{beenakker2013search}%
  \BibitemOpen
  \bibfield  {author} {\bibinfo {author} {\bibfnamefont {C.}~\bibnamefont {Beenakker}},\ }\href@noop {} {\bibfield  {journal} {\bibinfo  {journal} {Annu. Rev. Condens. Matter Phys.}\ }\textbf {\bibinfo {volume} {4}},\ \bibinfo {pages} {113} (\bibinfo {year} {2013})}\BibitemShut {NoStop}%
\bibitem [{\citenamefont {Elliott}\ and\ \citenamefont {Franz}(2015)}]{elliott2015colloquium}%
  \BibitemOpen
  \bibfield  {author} {\bibinfo {author} {\bibfnamefont {S.~R.}\ \bibnamefont {Elliott}}\ and\ \bibinfo {author} {\bibfnamefont {M.}~\bibnamefont {Franz}},\ }\href@noop {} {\bibfield  {journal} {\bibinfo  {journal} {Reviews of Modern Physics}\ }\textbf {\bibinfo {volume} {87}},\ \bibinfo {pages} {137} (\bibinfo {year} {2015})}\BibitemShut {NoStop}%
\bibitem [{\citenamefont {Lutchyn}\ \emph {et~al.}(2018)\citenamefont {Lutchyn}, \citenamefont {Bakkers}, \citenamefont {Kouwenhoven}, \citenamefont {Krogstrup}, \citenamefont {Marcus},\ and\ \citenamefont {Oreg}}]{lutchyn2018majorana}%
  \BibitemOpen
  \bibfield  {author} {\bibinfo {author} {\bibfnamefont {R.~M.}\ \bibnamefont {Lutchyn}}, \bibinfo {author} {\bibfnamefont {E.~P.}\ \bibnamefont {Bakkers}}, \bibinfo {author} {\bibfnamefont {L.~P.}\ \bibnamefont {Kouwenhoven}}, \bibinfo {author} {\bibfnamefont {P.}~\bibnamefont {Krogstrup}}, \bibinfo {author} {\bibfnamefont {C.~M.}\ \bibnamefont {Marcus}}, \ and\ \bibinfo {author} {\bibfnamefont {Y.}~\bibnamefont {Oreg}},\ }\href@noop {} {\bibfield  {journal} {\bibinfo  {journal} {Nature Reviews Materials}\ }\textbf {\bibinfo {volume} {3}},\ \bibinfo {pages} {52} (\bibinfo {year} {2018})}\BibitemShut {NoStop}%
\bibitem [{\citenamefont {Prada}\ \emph {et~al.}(2020)\citenamefont {Prada}, \citenamefont {San-Jose}, \citenamefont {de~Moor}, \citenamefont {Geresdi}, \citenamefont {Lee}, \citenamefont {Klinovaja}, \citenamefont {Loss}, \citenamefont {Nyg{\aa}rd}, \citenamefont {Aguado},\ and\ \citenamefont {Kouwenhoven}}]{prada2020andreev}%
  \BibitemOpen
  \bibfield  {author} {\bibinfo {author} {\bibfnamefont {E.}~\bibnamefont {Prada}}, \bibinfo {author} {\bibfnamefont {P.}~\bibnamefont {San-Jose}}, \bibinfo {author} {\bibfnamefont {M.~W.}\ \bibnamefont {de~Moor}}, \bibinfo {author} {\bibfnamefont {A.}~\bibnamefont {Geresdi}}, \bibinfo {author} {\bibfnamefont {E.~J.}\ \bibnamefont {Lee}}, \bibinfo {author} {\bibfnamefont {J.}~\bibnamefont {Klinovaja}}, \bibinfo {author} {\bibfnamefont {D.}~\bibnamefont {Loss}}, \bibinfo {author} {\bibfnamefont {J.}~\bibnamefont {Nyg{\aa}rd}}, \bibinfo {author} {\bibfnamefont {R.}~\bibnamefont {Aguado}}, \ and\ \bibinfo {author} {\bibfnamefont {L.~P.}\ \bibnamefont {Kouwenhoven}},\ }\href@noop {} {\bibfield  {journal} {\bibinfo  {journal} {Nature Reviews Physics}\ }\textbf {\bibinfo {volume} {2}},\ \bibinfo {pages} {575} (\bibinfo {year} {2020})}\BibitemShut {NoStop}%
\bibitem [{\citenamefont {Flensberg}\ \emph {et~al.}(2021)\citenamefont {Flensberg}, \citenamefont {von Oppen},\ and\ \citenamefont {Stern}}]{flensberg2021engineered}%
  \BibitemOpen
  \bibfield  {author} {\bibinfo {author} {\bibfnamefont {K.}~\bibnamefont {Flensberg}}, \bibinfo {author} {\bibfnamefont {F.}~\bibnamefont {von Oppen}}, \ and\ \bibinfo {author} {\bibfnamefont {A.}~\bibnamefont {Stern}},\ }\href@noop {} {\bibfield  {journal} {\bibinfo  {journal} {Nature Reviews Materials}\ }\textbf {\bibinfo {volume} {6}},\ \bibinfo {pages} {944} (\bibinfo {year} {2021})}\BibitemShut {NoStop}%
\bibitem [{\citenamefont {Yazdani}\ \emph {et~al.}(2023)\citenamefont {Yazdani}, \citenamefont {Von~Oppen}, \citenamefont {Halperin},\ and\ \citenamefont {Yacoby}}]{yazdani2023hunting}%
  \BibitemOpen
  \bibfield  {author} {\bibinfo {author} {\bibfnamefont {A.}~\bibnamefont {Yazdani}}, \bibinfo {author} {\bibfnamefont {F.}~\bibnamefont {Von~Oppen}}, \bibinfo {author} {\bibfnamefont {B.~I.}\ \bibnamefont {Halperin}}, \ and\ \bibinfo {author} {\bibfnamefont {A.}~\bibnamefont {Yacoby}},\ }\href@noop {} {\bibfield  {journal} {\bibinfo  {journal} {Science}\ }\textbf {\bibinfo {volume} {380}},\ \bibinfo {pages} {eade0850} (\bibinfo {year} {2023})}\BibitemShut {NoStop}%
\bibitem [{\citenamefont {Liu}\ \emph {et~al.}(2012)\citenamefont {Liu}, \citenamefont {Potter}, \citenamefont {Law},\ and\ \citenamefont {Lee}}]{liu2012zero}%
  \BibitemOpen
  \bibfield  {author} {\bibinfo {author} {\bibfnamefont {J.}~\bibnamefont {Liu}}, \bibinfo {author} {\bibfnamefont {A.~C.}\ \bibnamefont {Potter}}, \bibinfo {author} {\bibfnamefont {K.~T.}\ \bibnamefont {Law}}, \ and\ \bibinfo {author} {\bibfnamefont {P.~A.}\ \bibnamefont {Lee}},\ }\href@noop {} {\bibfield  {journal} {\bibinfo  {journal} {Physical review letters}\ }\textbf {\bibinfo {volume} {109}},\ \bibinfo {pages} {267002} (\bibinfo {year} {2012})}\BibitemShut {NoStop}%
\bibitem [{\citenamefont {Proke{\v{s}}}\ \emph {et~al.}(2020)\citenamefont {Proke{\v{s}}}, \citenamefont {Bartkowiak}, \citenamefont {Gorbunov}, \citenamefont {Prokhnenko}, \citenamefont {Rivin},\ and\ \citenamefont {Smeibidl}}]{prokevs2020noncollinear}%
  \BibitemOpen
  \bibfield  {author} {\bibinfo {author} {\bibfnamefont {K.}~\bibnamefont {Proke{\v{s}}}}, \bibinfo {author} {\bibfnamefont {M.}~\bibnamefont {Bartkowiak}}, \bibinfo {author} {\bibfnamefont {D.}~\bibnamefont {Gorbunov}}, \bibinfo {author} {\bibfnamefont {O.}~\bibnamefont {Prokhnenko}}, \bibinfo {author} {\bibfnamefont {O.}~\bibnamefont {Rivin}}, \ and\ \bibinfo {author} {\bibfnamefont {P.}~\bibnamefont {Smeibidl}},\ }\href@noop {} {\bibfield  {journal} {\bibinfo  {journal} {Physical Review Research}\ }\textbf {\bibinfo {volume} {2}},\ \bibinfo {pages} {013137} (\bibinfo {year} {2020})}\BibitemShut {NoStop}%
\bibitem [{\citenamefont {Shi}\ \emph {et~al.}(2021)\citenamefont {Shi}, \citenamefont {Parker}, \citenamefont {Choi}, \citenamefont {Devlin}, \citenamefont {Yin}, \citenamefont {Zhao}, \citenamefont {Klavins}, \citenamefont {Kauzlarich},\ and\ \citenamefont {Taufour}}]{shi2021robust}%
  \BibitemOpen
  \bibfield  {author} {\bibinfo {author} {\bibfnamefont {Y.}~\bibnamefont {Shi}}, \bibinfo {author} {\bibfnamefont {D.~S.}\ \bibnamefont {Parker}}, \bibinfo {author} {\bibfnamefont {E.~S.}\ \bibnamefont {Choi}}, \bibinfo {author} {\bibfnamefont {K.~P.}\ \bibnamefont {Devlin}}, \bibinfo {author} {\bibfnamefont {L.}~\bibnamefont {Yin}}, \bibinfo {author} {\bibfnamefont {J.}~\bibnamefont {Zhao}}, \bibinfo {author} {\bibfnamefont {P.}~\bibnamefont {Klavins}}, \bibinfo {author} {\bibfnamefont {S.~M.}\ \bibnamefont {Kauzlarich}}, \ and\ \bibinfo {author} {\bibfnamefont {V.}~\bibnamefont {Taufour}},\ }\href@noop {} {\bibfield  {journal} {\bibinfo  {journal} {Physical Review B}\ }\textbf {\bibinfo {volume} {104}},\ \bibinfo {pages} {184407} (\bibinfo {year} {2021})}\BibitemShut {NoStop}%
\bibitem [{\citenamefont {Wu}\ \emph {et~al.}(2019)\citenamefont {Wu}, \citenamefont {He}, \citenamefont {Han}, \citenamefont {Xu}, \citenamefont {Wu}, \citenamefont {Lin}, \citenamefont {Zhang}, \citenamefont {He},\ and\ \citenamefont {Wang}}]{wu2019induced}%
  \BibitemOpen
  \bibfield  {author} {\bibinfo {author} {\bibfnamefont {Y.}~\bibnamefont {Wu}}, \bibinfo {author} {\bibfnamefont {J.~J.}\ \bibnamefont {He}}, \bibinfo {author} {\bibfnamefont {T.}~\bibnamefont {Han}}, \bibinfo {author} {\bibfnamefont {S.}~\bibnamefont {Xu}}, \bibinfo {author} {\bibfnamefont {Z.}~\bibnamefont {Wu}}, \bibinfo {author} {\bibfnamefont {J.}~\bibnamefont {Lin}}, \bibinfo {author} {\bibfnamefont {T.}~\bibnamefont {Zhang}}, \bibinfo {author} {\bibfnamefont {Y.}~\bibnamefont {He}}, \ and\ \bibinfo {author} {\bibfnamefont {N.}~\bibnamefont {Wang}},\ }\href@noop {} {\bibfield  {journal} {\bibinfo  {journal} {Physical Review B}\ }\textbf {\bibinfo {volume} {99}},\ \bibinfo {pages} {121406} (\bibinfo {year} {2019})}\BibitemShut {NoStop}%
\end{thebibliography}%

\clearpage
		\onecolumngrid
\begin{center}
		\textbf{\large Supplemental Material for ``Pseudo-Ising superconductivity induced by $p$-wave magnetism''}\\[.2cm]
		 Zi-Ting Sun,$^{1}$  Xilin Feng,$^{1}$  Ying-Ming Xie,$^{2}$  Benjamin T. Zhou,$^{3}$  Jin-Xin Hu,$^{1}$  K. T. Law$^{1}$\\[.1cm]
		
        {\itshape ${}^1$Department of Physics, Hong Kong University of Science and Technology, Clear Water Bay, Hong Kong, China}
        {\itshape ${}^2$RIKEN Center for Emergent Matter Science (CEMS), Wako, Saitama 351-0198, Japan}
        
        {\itshape ${}^3$Thrust of Advanced Materials \& Quantum Science and Technology Center, The Hong Kong University of Science and Technology (Guangzhou), Nansha, Guangzhou, China}

\end{center}

\setcounter{equation}{0}
\setcounter{section}{0}
\setcounter{figure}{0}
\setcounter{table}{0}
\setcounter{page}{1}
\renewcommand{\theequation}{S\arabic{equation}}
\renewcommand{\thesection}{ \Roman{section}}

\renewcommand{\thefigure}{S\arabic{figure}}
\renewcommand{\thetable}{\arabic{table}}
\renewcommand{\tablename}{Supplementary Table}

\renewcommand{\bibnumfmt}[1]{[S#1]}
\makeatletter

\maketitle
\section{\bf{\uppercase\expandafter{I. Low-energy effective model of $p$-wave magnet}}}
In this section, we derive the low-energy effective Hamiltonian in the main text from the minimal tight-binding model introduced in~\cite{hellenes2023exchange,chakraborty2024highly}, which reads
\begin{equation}
H=2t(\cos k_y+\cos \frac{k_x}{2}\tau_x)-\mu+2t_J(\sin \frac{k_x}{2}\sigma_x \tau_y+\cos k_y\sigma_y \tau_z),
\label{eq:tbdmodel}
\end{equation}
where the Pauli matrices $\sigma$ and $\tau$ act on the spin and sublattice degrees of freedom, respectively. The first term is the spin-independent hopping term. The second term is the exchange-dependent hopping term, characterizing the $p$-wave spin splitting. We can project out the sublattice degree of freedom $\tau$ for the hole-like band around $\Gamma$ pocket [Fig.~\ref{fig:supfig1}~(a)]. In doing so, we treat $V=2t_J(\sin \frac{k_x}{2}\sigma_x \tau_y+\cos k_y\sigma_y \tau_z)$ as the perturbation term, and after the second-order perturbation theory (the first-order perturbation term is zero), we derive the effective two-band model as
\begin{equation}
H_{\mathrm{eff}}=2t\left(\cos k_y+\cos \frac{k_x}{2}\right)-\mu+\frac{t_J^2}{t \cos \frac{k_x}{2}}\left[\sin^2 \frac{k_x}{2}+\cos^2 k_y-2\sin \frac{k_x}{2} \cos k_y \sigma_z\right].
\end{equation}
In the continuum limit close to the band edge, we obtain an effective model containing only the spin degree of freedom near the $\Gamma$ pocket as 
\begin{equation}
   H_{eff}=-t_y k_y^2-t_x k_x^2-\mu'+J_p k_x(1-\frac{k_y^2}{2}+\frac{k_x^2}{8})\sigma_z, 
\end{equation}
where $t_y=t-\frac{J_p}{2}$, $t_x=\frac{t}{4}+\frac{3J_p}{8}$, $J_p=-\frac{t_J^2}{t}$, $\mu'=\mu-4t+J_p$. This continuum model can be discretized into a simpler lattice model as
\begin{equation}
   H_{eff} \approx 2t_y \cos k_y+2t_x \cos k_x-\mu''+J_p \sin k_x \cos k_y\sigma_z,
\end{equation}
where $\mu''=\mu-\frac{3t}{2}+\frac{3J_p}{4}$. In the main text, we use this simplified two-band tight-binding model to describe the $p$-wave magnetism.

\section{\bf{\uppercase\expandafter{II. Edelstein effect}}}

We provide more details in evaluating the superconducting magneto-electric effect. In the Nambu basis $\Psi_{\bm{k},\bm{q}}=(\hat{c}_{\bm{k}+\bm{q}/2,\uparrow},\hat{c}_{\bm{k}+\bm{q}/2,\downarrow},\hat{c}^\dagger_{-\bm{k}+\bm{q}/2,\uparrow},\hat{c}^\dagger_{-\bm{k}+\bm{q}/2,\downarrow})^T$, the finite-$\bm{q}$ mean field Hamiltonian is $H_{\mathrm{mf}}(\bm{q})=\frac{1}{2}\Psi_{\bm{k},\bm{q}}^\dagger H_{\mathrm{BdG}}(\boldsymbol{k},\bm{q})\Psi_{\bm{k},\bm{q}}$. The Bogoliubov-de Gennes (BdG) Hamiltonian reads
\begin{equation}\label{eq:bdg}
H_{\mathrm{BdG}}(\boldsymbol{k},\bm{q})=\left(\begin{array}{cc}
h(\boldsymbol{k}+\bm{q}/2) & -i \Delta \sigma_y \\
i \Delta \sigma_y & -h^*(-\boldsymbol{k}-\bm{q}/2)
\end{array}\right).
\end{equation}
where $h(\bm{k})$ is the Bloch Hamiltonian of the normal state, $\bm{q}$ is the momentum of the Cooper pair, and $s_y$ is a Pauli matrix. For simplicity, we have used a $\bm{q}$-independent gap function, which is valid for weak values of the applied current. The supercurrent $J_s=\nu_s \bm{q}$ with the superfluid density $\nu_s=\partial^2 \mathcal{F}_{\mathrm{sc}}(\bm{q})/\partial^2\bm{q}$. 

\begin{figure}[t]
		\centering
		\includegraphics[width=0.8\linewidth]{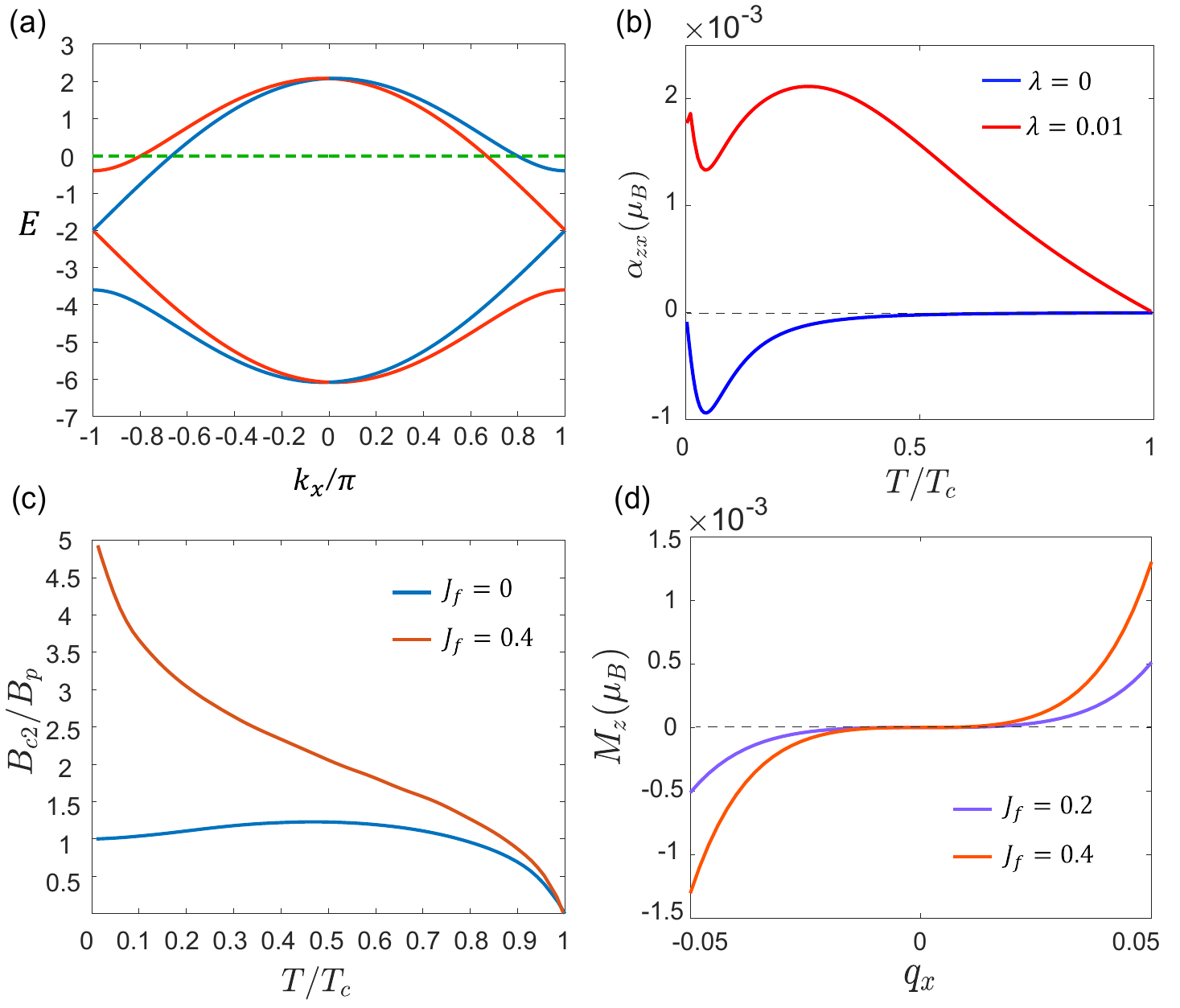}
		\caption{(a) Band structure from Eq.~\eqref{eq:tbdmodel} with parameters $(t,t_J, \mu,k_y) = (2, 0.4,6,0)$. (b) Edelstein effect in $p$-wave magnet with $(J_p,\Delta_0, \mu) = (0.3, 0.06,-1)$. (c) $B_{c2}\text{--}T$ curve of $f$-wave magnet with $(t, \mu) = (2, 2)$. (d) Edelstein effect in $f$-wave magnet with $t=1,\mu=0.3,\lambda=0.05,\Delta_0=0.05, T=0.3T_c$. We add the SOC term as $H_{soc}=\lambda(k_y \sigma_x-k_x \sigma_y)$.}
		\label{fig:supfig1}
\end{figure}

In the current-carrying state, the magnetization $M$ can be written as
\begin{equation}
\label{eq:magnetization}
M_a=-\frac{g_s\mu_B}{4\beta}\sum_{n\bm{k}}\mathrm{Tr}[\mathcal{G}(\bm{k},\bm{q},i\omega_n)\eta_a].
\end{equation}
Here $\eta_a=\mathrm{diag}(\sigma_a,-\sigma_a^*)$ is the redefined spin Pauli matrices in the Nambu space. The Gor'kov Green's function is $\mathcal{G}(\bm{k},\bm{q},i\omega_n)=[i\omega_n-H_{\mathrm{BdG}}(\boldsymbol{k},\bm{q})]^{-1}$ with the Matsubara frequency $\omega_n=(2n+1)\pi k_B T$. $T$ is the temperature. $g_s=2$ is the Land\'{e} $g$ factor and $\mu_B$ is the Bohr magneton. To derive the linear Edelstein effect, we can then expand the BdG Hamiltonian as 
\begin{equation}
\label{eq:HBdG}
H_{\mathrm{BdG}}(\boldsymbol{k},\bm{q})=H_{\mathrm{BdG}}(\boldsymbol{k},0)+q_a \hat{v}_a /2+\mathcal{O}(\bm{q}^2),
\end{equation}
where the velocity operator is 
\begin{equation}
\hat{v}_a=\left(\begin{array}{cc}
 \hat{\mathcal{V}}_{a}(\bm{k})& 0 \\
0 & -\hat{\mathcal{V}}^*_{a}(-\bm{k})
\end{array}\right),
\end{equation}
Here we introduce $\mathcal{V}_a(\bm{k})=\partial H_{\bm{k}}/\partial k_a$. Then we obtain
\begin{equation}
\label{eq:green_expand}
\mathcal{G}(\bm{k},\bm{q},i\omega_n)=[i\omega_n-H_{\mathrm{BdG}}(\boldsymbol{k},\bm{q})]^{-1}=\mathcal{G}_0+\frac{1}{2}q_a \mathcal{G}_0 \hat{v}_a \mathcal{G}_0
\end{equation}
Here $\mathcal{G}_0 \equiv \mathcal{G}(\bm{k},i\omega_n)=(i\omega_n-H_{\mathrm{BdG}}(\boldsymbol{k},0))^{-1}$. We insert Eq.~\eqref{eq:green_expand} to Eq.~\eqref{eq:magnetization} to get the magneto-electric susceptibility $\alpha_{ab}$ reads
\begin{equation}
\label{eq:linear_sus}
\alpha_{ab}=-\frac{g_s\mu_B k_B T}{8}\sum_{n\bm{k}}\mathrm{Tr}[\eta_a \mathcal{G}_0 \hat{v}_b \mathcal{G}_0].
\end{equation}
where the magnetization is described by $\delta M_a=\alpha_{ab} q_b$. Here we compare $\alpha_{zx}$ in $p$-wave magnet with and without a small Rashba SOC in Fig.~\ref{fig:supfig1}~(b) as a supplement to the main text.

\section{\bf{\uppercase\expandafter{III. Pseudo-Ising superconductivity with odd-parity $f$-wave magnetism}}}
Apart from $p$-wave magnetism, we show that the pseudo-Ising superconductivity is a general feature for odd-parity magnetism. To demonstrate this, we take $f$-wave magnetism with $\gamma_f(\boldsymbol{k})=J_fk_x(k_x^2-3k_y^2)$ type nonrelativistic spin-momentum locking as an example. According to Eq.~\eqref{eq:paircorrelation} in the main text, if we replace $\gamma_p(\boldsymbol{k})$ by $\gamma_f(\boldsymbol{k})$, the presence of $f$-wave magnetism produces a nonvanishing out-of-plane $f$-wave spin-triplet pairing correlation (in-plane equal-spin pairing correlation), which is $\propto \gamma_f(\boldsymbol{k})\hat{z}$ despite the mean-field pairing potential being $s$-wave.

Combining Eq.~\eqref{eq:paircorrelation} and Eq.~\eqref{eq:phi}, the self-consistent gap equation with $f$-wave magnetism under an in-plane magnetic field is given by \cite{xie2020strongly}:
\begin{equation}
1=T \sum_{k, n} U F_s\left(\boldsymbol{k}, V_x, i \omega_n, \Delta\right),   
\end{equation}
where $U$ is the coupling constant. More explicitly, we obtain
\begin{equation}
1=\frac{T U}{2} \sum_{\boldsymbol{k}, n}\left[1-\frac{V_x^2}{\lambda(\boldsymbol{k},\Delta,V_x)}\right] \frac{1}{\omega_n^2+\chi_{-}^2(\boldsymbol{k},\Delta,V_x)}+\left[1+\frac{V_x^2}{\lambda(\boldsymbol{k},\Delta,V_x)}\right] \frac{1}{\omega_n^2+\chi_{+}^2(\boldsymbol{k},\Delta,V_x)},
\end{equation}
where $\chi_{ \pm}(\boldsymbol{k},\Delta,V_x)=\left[V_x^2+\gamma_f^2(\boldsymbol{k})+\Delta^2+\xi^2(\boldsymbol{k}) \pm 2 \lambda(\boldsymbol{k},\Delta,V_x)\right]^{1 / 2}$, $\lambda(\boldsymbol{k},\Delta,V_x)=\left\{V_x^2\left[\Delta^2+\xi^2(\boldsymbol{k})\right]+\gamma_f^2(\boldsymbol{k}) \xi^2(\boldsymbol{k})\right\}^{1 / 2}$. Then we sum over the Matsubara frequencies and get
\begin{equation}
1=\frac{U}{2} \sum_{\boldsymbol{k}}\left[1-\frac{V_x^2}{\lambda(\boldsymbol{k},\Delta,V_x)}\right] \frac{1}{2 \chi_{-}(\boldsymbol{k},\Delta,V_x)} \tanh \left[\frac{\chi_{-}(\boldsymbol{k},\Delta,V_x)}{2 T}\right]+\left[1+\frac{V_x^2}{\lambda(\boldsymbol{k},\Delta,V_x)}\right] \frac{1}{2 \chi_{+}(\boldsymbol{k},\Delta,V_x)} \tanh \left[\frac{\chi_{+}(\boldsymbol{k},\Delta,V_x)}{2 T}\right] .
\label{eq:gapeq}
\end{equation}

To obtain $B_{c2}\text{--}T$ relation, we set $\Delta=0$ and $V_x=V_{c2}$ in Eq.~\eqref{eq:gapeq} and reduce it to the linearized gap equation: 
\begin{equation}
\frac{4}{U}=\sum_{\boldsymbol{k}} \left[1+\frac{V_{c2}^2}{\sqrt{V_{c2}^2+\gamma_f^2} |\xi|}\right]\frac{\tanh \left[\frac{|\xi|+\sqrt{V_{c2}^2+\gamma_f^2}}{2 T}\right]}{|\xi|+\sqrt{V_{c2}^2+\gamma_f^2}}+\left[1-\frac{V_{c2}^2}{\sqrt{V_{c2}^2+\gamma_f^2} |\xi|}\right]\frac{\tanh \left[\frac{||\xi|-\sqrt{V_{c2}^2+\gamma_f^2}|}{2 T}\right]}{||\xi|-\sqrt{V_{c2}^2+\gamma_f^2}|}.
\label{eq:ligapeq1}
\end{equation}
When $V_{c2}=0$, $T=T_c$, and we have
\begin{equation}
\frac{4}{U}=\sum_{\boldsymbol{k}} \frac{\tanh \left[\frac{|\xi(\boldsymbol{k})|+|\gamma_f(\boldsymbol{k})|}{2 T_c}\right]}{|\xi(\boldsymbol{k})|+|\gamma_f(\boldsymbol{k})|}+\frac{\tanh \left[\frac{||\xi(\boldsymbol{k})|-|\gamma_f(\boldsymbol{k})||}{2 T_c}\right]}{||\xi(\boldsymbol{k})|-|\gamma_f(\boldsymbol{k})||} .
\label{eq:ligapeq2}
\end{equation}
Replacing the left hand side of Eq.~\eqref{eq:ligapeq1} by the right hand side of Eq.~\eqref{eq:ligapeq2}, we can plot $B_{c2}\text{--}T$ curve in Fig.~\ref{fig:supfig1}~(c), with $\xi(\boldsymbol{k})=t(k_x^2+k_y^2)-\mu$. And the $B_{c2}$ is strongly enhanced in the presence of $f$-wave magnetism, similar to the case of $p$-wave magnetism.

Here, it is also intriguing to consider the vertical Edelstein effect. In this case, we directly evaluate the current-induced magnetization from Eq.~\eqref{eq:magnetization}, and the results are presented in Fig.~\ref{fig:supfig1}~(d). The leading order contribution of $M_z$ is proportional to $q_x^{3}$.

\begin{figure}[t]
		\centering
		\includegraphics[width=0.8\linewidth]{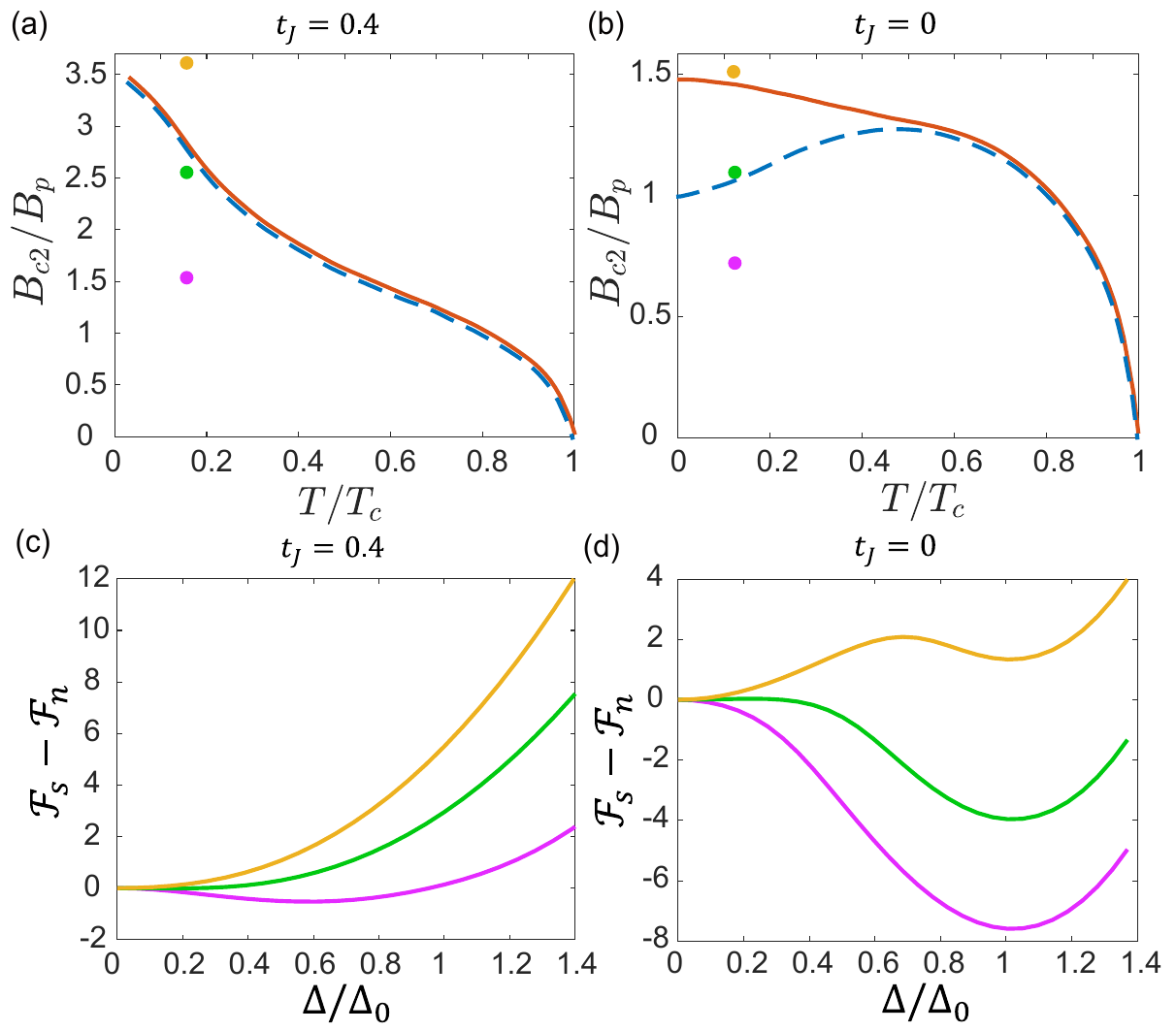}
		\caption{(a) and (b): $B_{c2}\text{--}T$ for Eq.~\eqref{eq:tbdmodel}, with $(t, \mu,U) = (2,6,2)$. The dashed lines are from the linearized gap equations, while the solid lines are from minimizing $\mathcal{F}_s$. (c) and (d): The free energy density profiles with respect to pairing amplitude $\Delta$. The curves in (c) and (d) correspond to the highlight points in (a) and (b), respectively.}
		\label{fig:supfig2}
\end{figure}

\section{\bf{\uppercase\expandafter{IV. Calculations for four-band models}}}

In the main text, we have studied the pseudo-Ising superconductivity based on a simple two-band model to highlight the role of $p$-wave magnetism. Note that the main physics does not rely on the specific model Hamiltonian. Here we examine both Eq.~\eqref{eq:tbdmodel} and another four-band tight-binding model with $\mathcal{T}\bm{\tau}$ symmetry, which has been proposed recently~\cite{brekke2024minimal} and reads
\begin{equation}
h_0(\bm{k})=-\{ 2t[\cos(k_x)+\cos(k_y)]+\mu\}\sigma_0 \otimes \tau_0+[\alpha_x \sin(k_x)+\alpha_y \sin(k_y)]\sigma_z \otimes \tau_0+J_{sd}\sigma_x \otimes \tau_z,
\label{eq:linder}
\end{equation}
where $J_{sd}$ denotes the isotropic $s$-$d$ coupling between the localized spin and itinerant electrons. $\alpha_x,\alpha_y$ are spin-dependent hoppings to reveal the noncollinear magnetism. We introduce the model Hamiltonian of the superconductor with $p$-wave magnet via the inverse proximity effect as
\begin{equation}
H=  \sum_{\boldsymbol{k}, \alpha, \beta, s, s^{\prime}} c_{\boldsymbol{k}, \alpha, s}^{\dagger}h(\boldsymbol{k})_{\alpha s, \beta s^{\prime}} c_{\boldsymbol{k}, \beta, s^{\prime}} -\frac{U}{A} \sum_{\boldsymbol{k}, \boldsymbol{k}^{\prime}} \sum_{\alpha, \beta} c_{\boldsymbol{k}, \alpha, \uparrow}^{\dagger} c_{-\boldsymbol{k}, \alpha, \downarrow}^{\dagger} c_{-\boldsymbol{k}^{\prime}, \beta, \downarrow} c_{\boldsymbol{k}^{\prime}, \beta, \uparrow}
\label{eq:pairing}
\end{equation}
where $c_{\boldsymbol{k}, \alpha, s}^{\dagger}$ is the electronic creation operator, $\alpha$ is the orbital (sublattice) index, and $s=\uparrow / \downarrow$ represents the spin index. $U$ is the coupling constant of the intra-sublattice attractive interaction, and $A$ is the area of the interface. The hopping matrix $h(\boldsymbol{k})$ can be written as $h(\boldsymbol{k})=h_0(\boldsymbol{k})+h_Z$ with the Zeeman term $h_Z=\frac{1}{2} g_s \mu_{\mathrm{B}} \boldsymbol{B} \cdot (\boldsymbol{\sigma}\otimes \tau_0)$, $\boldsymbol{B}=(B_x,B_y,0)$ representing the in-plane magnetic field. 

Here we want to illustrate more about the continuous nature of the phase transition near $B_{c2}$ for the pseudo-Ising superconductor in terms of the model Eq.~\eqref{eq:tbdmodel} and Eq.~\eqref{eq:pairing}. We use two different ways to calculate the $B_{c2}$, the first one is by the linearized gap equation (Eq.~\eqref{lineareq} in the main text), and the second one is to directly minimize the free energy density $\mathcal{F}_s$ (Eq.~\eqref{freen} in the main text) and see when the global minimum will become $\Delta=0$. We can define $\mathcal{F}_n=\mathcal{F}_s(\Delta=0)$ for simplicity. The results are summarized in Fig.~\ref{fig:supfig2}. Comparing (a) and (b), we can observe that the existence of $p$-magnetism strongly enhances the $B_{c2}$. And for $t_J=0.4$ (Fig.~\ref{fig:supfig2}~(a)), the $B_{c2}\text{--}T$ curves from the two methods coincide with each other, while they do not for $t_J=0$ (Fig.~\ref{fig:supfig2}~(b)). To understand it, we can further compare (c) and (d). (c) implies that the phase transition at $B_{c2}$ in (a) is second-order, and (d) implies that the real superconductor-metal phase transition for $t_J=0$ happens at the solid line rather than the dashed line, which is a first-order phase transition. So for the former case, the linearized gap equation can be used to indicate the transition point when the global minimum of $\mathcal{F}_s$ becomes $\Delta=0$. But for the latter case, the linearized gap equation only implies that $\Delta=0$ becomes a local minimum of $\mathcal{F}_s$, while the global minimum of $\mathcal{F}_s$ can still be at $\Delta\neq0$.

\begin{figure}[t]
		\centering
		\includegraphics[width=0.8\linewidth]{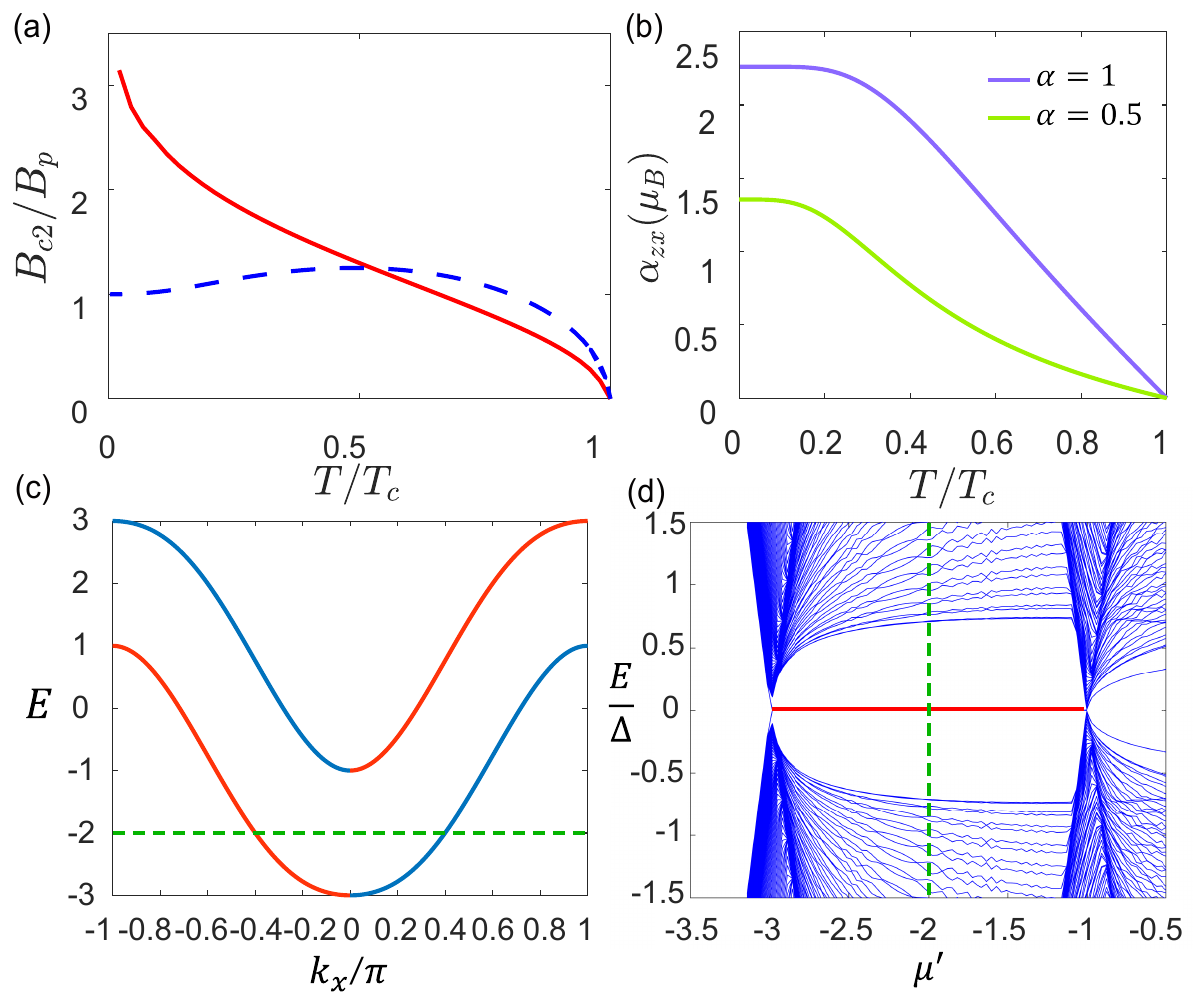}
		\caption{(a) $B_{c2}\text{--}T$ for the four-band model Eq.~\eqref{eq:linder} with $t=1$, $\alpha=J_{sd}=1$, $U=2.5$ and $\mu=-4$. For comparison, we plot the conventional case with $t=1$, $\alpha=J_{sd}=0$, $U=2.5$ and $\mu=-3$. (b) Edelstein effect with $t=1$, $J_{sd}=1$, $\Delta_0=0.1$ for $\alpha=1$ and $0.5$. (c) Band structure from Eq.~\eqref{eq:linder} with parameters $(t,\alpha,J_{sd},k_y) = (1, 1,1,0)$, $\mu=-2t$. (d) The BdG spectrum of the superconducting QW made of Eq.~\eqref{eq:linder} as a function of the effective chemical potential $\mu'=\mu+2t$ of the wire. The red line highlights the topological regime with MZMs. Here we only plot the $\Delta^2+(2t+\mu')^2<J_{sd}^2$ branch as a representative. Parameters: $\alpha=1$, $t=1$, $\Delta=0.1$, $J_{sd}=1$.}
		\label{fig:supfig3}
\end{figure}

Then we move to model Eq.~\eqref{eq:linder}. The results of $B_{c2}$ and linear Edelstein effect are summarized in Fig.~\ref{fig:supfig3}~(a) and (b). The conservation of $\langle s_z \rangle$ in Eq.~\eqref{eq:linder} is broken by $s$-$d$ coupling, so that the dominant out-of-plane linear Edelstein effect can emerge without Rashba spin-orbit coupling. Interestingly, the model Eq.~\eqref{eq:linder} has two unmixed sectors, and for each sector, the $s$-$d$ coupling plays the role of an internal in-plane magnetic field, but has opposite signs between these two sectors. If we restrict this system to a one-dimensional wire by setting $k_y=0$ and taking the superconducting pairing as in Eq.~\eqref{eq:pairing}, each sector of this model can be exactly described by Eq.~\eqref{majorana} in the main text. So the MZMs can emerge separately for each sector in this model, even without external magnetic fields (of course, it is robust against external magnetic fields), as long as the two sectors do not couple with each other. Similar to the case introduced in the main text, the topological regime is given by $\Delta^2+(2t+\mu')^2<J_{sd}^2$. The band structure and topological regime are plotted in Fig.~\ref{fig:supfig3}~(c) and (d), respectively.

\end{document}